\documentclass[aps,pra,twocolumn,showpacs,superscriptaddress,nofootinbib]{revtex4-1}   % for review and submission showpacs,
\usepackage{graphicx}  % needed for figures
\usepackage[caption=false]{subfig}
\usepackage{float}
\usepackage[toc,page]{appendix}
\usepackage[table]{xcolor}
\usepackage{dcolumn}   % needed for some tables
\usepackage{bm}        % for math
\usepackage{amssymb}   % for math
\usepackage{color}

\usepackage{hyperref}
\usepackage{pifont}
\usepackage{amsmath, amsthm, amssymb}
\usepackage{amsfonts}
\usepackage{relsize}    % to resize the sum symbol etc
\usepackage{bbold}       % for identity in tensor notation
\usepackage{accents}    % \underaccent { accent } { symbol }

\hypersetup{
    colorlinks=true, %set true if you want colored links
    linktoc=all,     %set to all if you want both sections and subsections linked
    linkcolor=blue,  %choose some color if you want links to stand out
    citecolor=blue,
    filecolor=blue,
    urlcolor=blue
}
\usepackage{soul}
\newcommand{\beq}{\begin{equation}}
\newcommand{\eeq}{\end{equation}}
\newcommand{\bqa}{\begin{eqnarray}}
\newcommand{\eqa}{\end{eqnarray}}
\newcommand{\nn}{\nonumber}

\newcommand{\erf}[1]{Eq.~(\ref{#1})}

\newcommand{\erfs}[2]{Eqs.~(\ref{#1})--(\ref{#2})}
\newcommand{\erfa}[2]{Eqs.~(\ref{#1}) and (\ref{#2})}
\newcommand{\arf}[1]{{App.}~\ref{#1}} 
\newcommand{\srf}[1]{Sec.~\ref{#1}} 
\newcommand{\crf}[1]{Ref.~\cite{#1}}

\newcommand{\frf}[1]{Fig.~\ref{#1}}
\newcommand{\fsrf}[2]{Fig.~\ref{#1}\subref{#2}}
\newcommand{\ea}{{\it et al.}}
\newcommand{\ie}{{\it i.e.}}
\newcommand{\dg}{^\dagger}

\definecolor{BLACK}{gray}{0}
\definecolor{RED}{rgb}{1,0,0}
\definecolor{GREEN}{rgb}{0.2,.6,0.2}
\definecolor{amber}{rgb}{1.0,0.49,0.0}
\newcommand{\pu}{\color{purple}}

\newcommand{\blk}{\color{BLACK}}

\renewcommand{\(}{\left(}
\renewcommand{\)}{\right)}
\renewcommand{\[}{\left[}
\renewcommand{\]}{\right]}
\newcommand{\sq}[1]{\left[ {#1} \right]}

\newcommand{\an}[1]{\left\langle{#1}\right\rangle}

\newcommand{\abs}[1]{\left| {#1} \right|}
\newcommand{\tr}[1]{{\rm Tr}\sq{ {#1} }}

\newcommand{\smallfrac}[2]{\mbox{$\frac{#1}{#2}$}}

\newcommand{\bra}[1]{\langle{#1}|}

\newcommand{\ket}[1]{|{#1}\rangle}
\newcommand{\bket}[1]{\Big|{#1}\Big\rangle}  
 
\newcommand{\op}[2]{\ket{#1}\bra{#2}}

\begin{document}

% The following information is for internal review, please remove them for submission
\widetext

% the following line is for submission, including submission to the arXiv!!
%\hspace{5.2in} \mbox{Fermilab-Pub-04/xxx-E}

\title{Experimental investigation of a multi-photon Heisenberg-limited interferometric  scheme: the effect of imperfections}

\author{Shakib Daryanoosh} 
\email{shakib.daryanoosh@uwa.edu.au}
\author{Geoff J. Pryde} 
\author{Howard M.~Wiseman} \email{h.wiseman@griffith.edu.au}
\author{Sergei Slussarenko} \email{s.slussarenko@griffith.edu.au}
\affiliation{Centre for Quantum Dynamics and Centre for Quantum Computation and Communication Technology, Griffith University, Yuggera Country, Brisbane,  Queensland 4111, Australia}

%
% visitor_addresses.tex                       26 March 2013 (a) 
%  available symbols are:
%  $\ast, \dag, \ddag, \S, \P, $\|$, $\ast\ast$, \dag\dag, \ddag\ddag ,\#
%

%{falkowski}
%$^{?}$Laboratoire de Physique Theorique, Orsay, FR,
%{hooper,kozminski}
%$^{?}$}Visitor from Lewis University, Romeoville, IL, USA.
%{weber}
%$^{?}$Universit{\"a}t Bern, Bern, Switzerland.
%{deceased}
%$^{\ddag}$Deceased.

%
\vskip 0.25cm
       % D0 authors (remove the first 3 lines
                             % of this file prior to submission, they
                             % contain a time stamp for the authorlist)
                             % (includes institutions and visitors)
\date{\today}

\begin{abstract} 
Interferometric phase estimation is an essential tool for precise measurements of quantities such as displacement, velocity and material properties. The lower bound on measurement uncertainty achievable with classical resources is set by the shot-noise limit (SNL) that scales asymptotically  as $1/\sqrt{N}$, where $N$ is the number of resources used. The experiment of S. Daryanoosh \ea\ [{\it Nat. Commun.} {\bf 9}, 4606 (2018)] showed how to achieve the ultimate precision limit, the exact Heisenberg limit (HL), in {\it ab-initio} phase estimation with $N=3$ photon-passes, using an entangled biphoton state in combination with particular measurement techniques. The advantage of the HL over the SNL increases with the number of resources used. Here we present, and implement experimentally, a scheme for generation of the optimal $N=7$ triphoton state. We study experimentally and theoretically the generated state quality and its potential for phase estimation. We show that the expected usefulness of the prepared triphoton state for HL phase estimation is significantly degraded
%penalty from ..very 
by even quite small experimental imperfections, such as optical mode mismatch and unwanted higher-order multi-photon terms in the states produced in parametric down-conversion. 
%pairs production in photon sources.  
\end{abstract} 

\maketitle

% ============================================
%	     			Introduction
% ============================================ 
\section{Introduction}
Phase measurement is an indispensable part of science and technology~\cite{WisMil10, Jac14} as it offers simple yet robust methods for measuring a variety of physical quantities. Quantum mechanics bounds the ultimate precision in measurements to the Heisenberg limit (HL), which scales reciprocally with ${N}$ (for large $N$), where $N$ is the number of quantum resources. This contrasts with the Shot Noise Limit (SNL), which is ${1}/{\sqrt{N}}$ asymptotically. Improving measurement precision beyond the SNL~\cite{GioMac11, PirLlo2018,MigBie13}, towards the goal of achieving the best precision possible, requires employing intrinsic quantum properties such as quantum superposition and entanglement in conjunction with other techniques. 

Unlike phase sensing~\cite{NagTak07,Dow08,YonFur12,SluPry17,thekkadath20,YouGerr21}, which deals with the maximum sensitivity achievable in a measurement of a small variation of an already known parameter, \emph{ab-initio} phase estimation~\cite{HigPry07, XiaPry10, BerAnd15, HasMac17, DarPry18} aims at determining the exact value of the phase with no prior knowledge. In this situation, the exact optimal quantum precision, the exact HL, is $\pi/N$ in the asymptotic regime~\cite{Gorecki20}.
 
This exact optimal precision can be achieved by an interferometric Heisenberg-limited phase estimation algorithm~\cite{WisPry09} (HPEA), as has since been demonstrated experimentally with photonic qubits~\cite{DarPry18}. That experiment relied on three techniques to attain uncertainty very close to the exact HL for $N=3$ resources: the use of the optimal entangled two-qubit state preparation, multiple application of the phase shift, and performing adaptive measurements~\cite{WisPry09}. Here, as is standard in quantum metrology~\cite{GioMac11}, a single resource corresponds to a single qubit passing through a rotation by the unknown phase. For the case $N=3$, one photonic qubit passes the phase shift once and the other twice.

Two-photon probe states are relatively easy to generate with spontaneous parametric downconversion (SPDC) photon sources and small-scale optical circuits. But to obtain more quantum advantage over classical measurement schemes, per resource used, larger, multi-photon probe states and thus more complex optical state generation schemes are needed. 

In this work we study both theoretically and experimentally, the $N=7$ version of HPEA in the presence of experimental imperfections.  We theoretically investigate the protocol under the influence of optical mode mismatch and noise coming from the unwanted photon emission events from the photon sources. We analyze these effects separately due to the computational complexity of simulations. Our results indicate that the quality of phase estimation is highly sensitive to these imperfections. Next we implement experimentally a setup for generating the optimal~\cite{WisPry09} three-qubit ($N=7$ resources) probe state. This is an unusual 3-photon state that as not been reported in prior experimental literature.  Feeding the tomographically reconstructed experimental probe state into a stochastic simulation of the phase estimation protocol, we find out that its quality is insufficient for measuring phase with precision near the HL. In fact, the phase uncertainty is greater than the SNL because of experimental challenges such as maintaining the setup stability over the course of experiment. 

The paper is outlined as the following. In \srf{sec:theory} we overview the Heisenberg-limited phase estimation algorithm and design a quantum circuit for creating the optimal state using three photonic polarization qubits. 
% in \srf{sec:mmc}
\srf{sec:imperfect} deals with modeling optical mode mismatch and multi-photon pairs generation in SPDC process, respectively. Experimental results are discussed in \srf{sec:exp}, followed by conclusion in \srf{sec:results}.

% ============================================
%	     				 Theory
% ============================================ 
\section{Theory} \label{sec:theory}

% -------------------------------------------------------------------------
%	      Interferometric phase estimation algorithm
% -------------------------------------------------------------------------
\subsection{Interferometric phase estimation}  \label{sec:IPEA}
HPEA can be executed on a modified Mach-Zehnder interferometer (MZI) depicted in~\frf{fig:GMZI}(a). The unknown phase  $\phi$ is placed in arm ``I'' of the interferometer. The arm is configured in such a way that it allows an optical mode to pass through a phase shift element $p$ times so that the total phase of $p \phi$ would be acquired by a single photon.  The other arm (``II'') of the interferometer contains a reference phase $\theta$. This phase can be adjusted in the course of the experiment~\cite{BerWis00, BerBre01}. That is, after each detection, the measurement outcomes are analyzed in a processor unit, and based on the results, $\theta$ is adjusted according to the protocol in~\crf{Wis95}. 

 % -----------------------------------------------------------------------------------
%					Figure
% -----------------------------------------------------------------------------------
\begin{figure}
	\captionsetup[subfigure]{labelformat=empty}
	\includegraphics[scale=0.27]{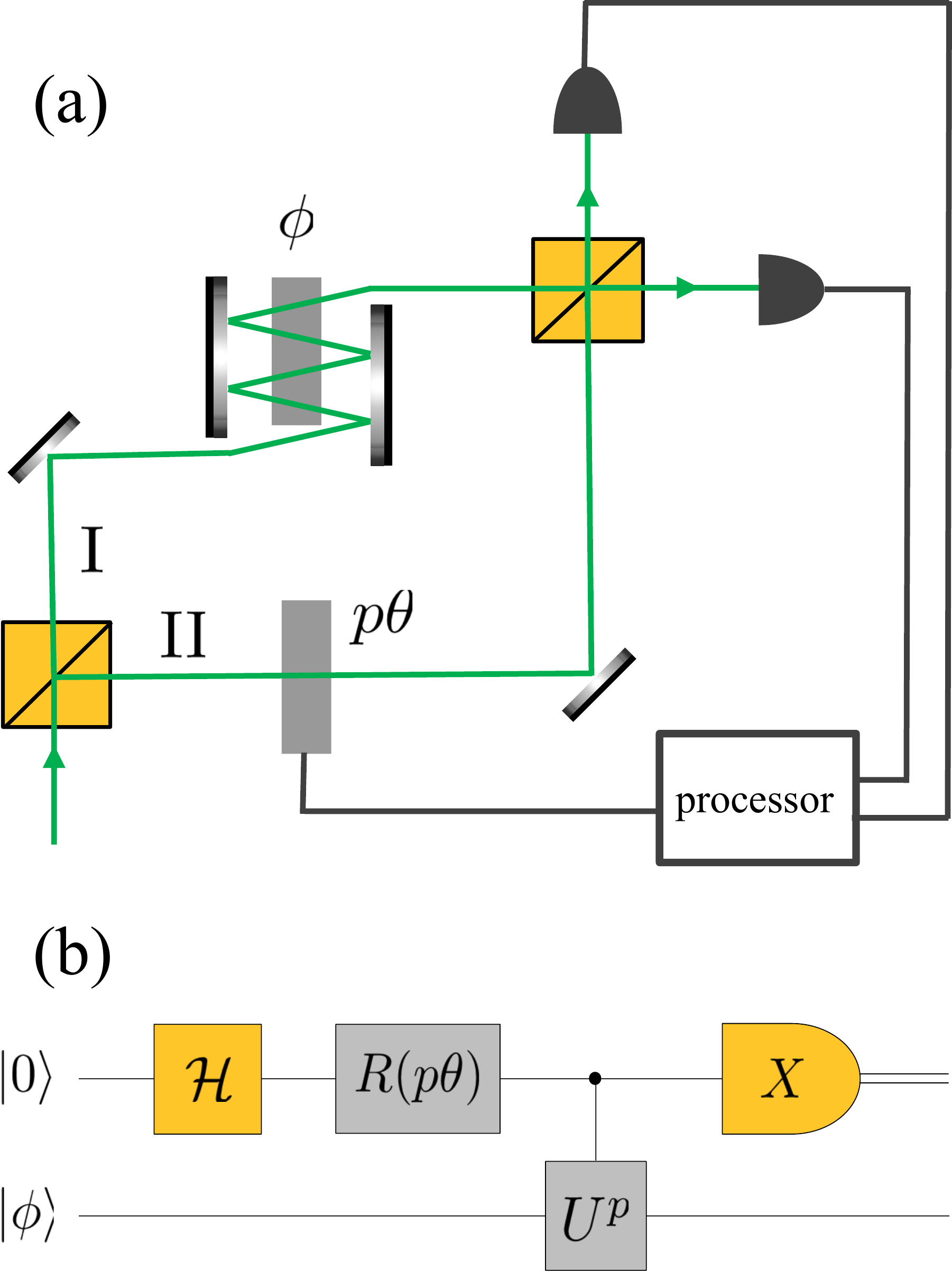}
	\caption{\label{fig:GMZI} (Color online) (a) Schematic representation of the modified Mach-Zehnder interferometer allowing for multiple application of the phase element $\phi$. The optical mode in path ``I" passes multiple ($p$) times (here $p = 4$) such that a total phase shift of $p\phi$ is acquired. Also the reference phase in path ``II" is set so that it imparts $\theta$ phase shift. (b) Quantum circuit illustration for the interferometric phase estimation scheme with one qubit using the interferometer from (a).}
\end{figure}

In the context of quantum information, an interferometric phase measurement scheme can be mapped on to a quantum circuit. This becomes handy when it comes to considering various phase estimation algorithms.
The two arms of the interferometer correspond to the two basis states  $\ket{0}$ and  $\ket{1}$ of a qubit. An input qubit is then represented by the logical state $\ket{0}$. A beam splitter (BS) that the photon impinges on performs a Hadamard operation ${\cal H} \ket{0} (\ket{1}) = \ket{+} (\ket{-})$, where $\ket{\pm} \equiv (\ket{0} \pm \ket{1})/\sqrt{2}$. The unknown phase shift $p \phi$ is represented by the controlled unitary (CU) operator $\op{1}{1} \otimes \hat U^p + \op{0}{0} \otimes \hat{I}$, where $\hat U \ket{\phi} = e^{i \phi} \ket{\phi}$ acts on an additional register in state $\ket{\phi}$ 
and $\hat I$ is the identity operator. The auxiliary phase shift $\theta$ is represented by $\hat R(\theta) \equiv {\rm exp}(i \theta \op{0}{0})$. The system state after the second BS undergoes basis transformation $Z \rightarrow X$, that is, from the logical to $\ket{\pm}$ basis. This implies that the latter BS can be replaced with a measurement stage in the $X$ basis. Therefore, estimating the phase using this interferometric algorithm is described by the quantum circuit illustrated in \frf{fig:GMZI}(b). 

Utilising path-entangled NOON states or multipassing can improve the sensitivity in a phase measurement compared to a repeated single-photon single-pass measurement~\cite{WisPry09}. At the same time they reduce the range of phase value that can be distinguished, making it impossible to distinguish phase values that differ by more than $\pi/p$. To perform an \emph{ab-initio} phase estimation one needs to remove this ambiguity in phase measurement and extend the  range of the measurement to the full  $[0, 2\pi)$ interval. This can be achieved with an appropriate choice of the probe state and measurement protocol.

% -----------------------------------------------------------------------------------
%					Figure 
% -----------------------------------------------------------------------------------
\begin{figure*}
	\includegraphics[scale=0.75]{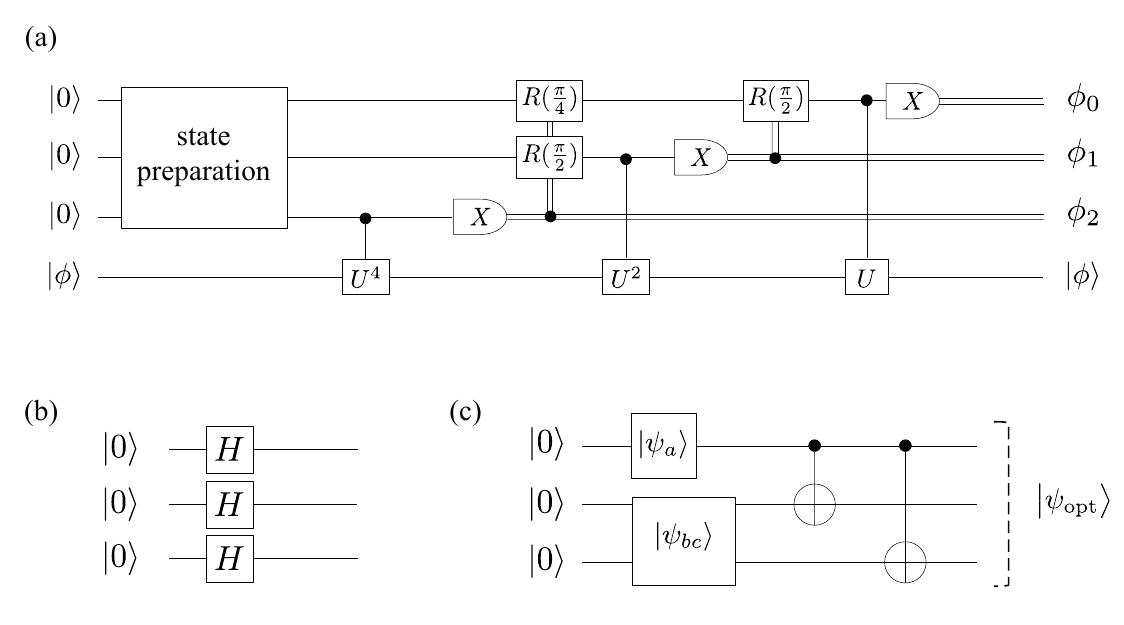}
 \caption{\label{fig:3photab} (a) Circuit diagram for the Heisenberg-limited interferometric phase estimation binary encoding for probe qubits with $N = 7$ quantum resources. (b) State preparation for the QPEA: a Hadamard operation is applied on each qubit in the same way depicted in \frf{fig:GMZI}. (c) Circuit representation for creating the 3-photon optimal state, \erf{s:opt:3phot}. The input state,$\ket{\psi_{\rm in}} = \ket{\psi_{a}} \ket{\psi_{bc}}$, is transformed into the optimal state, after application of two consecutive CNOT gates.} 
\end{figure*} 

In order to evaluate the performance of phase measurement protocols, we use the Holevo measure of deviation defined as \cite{Hol84} 
\beq    \label{eqn:holdev}
D_{\rm H} = \bigg|\Big\langle\big\langle{{\rm exp}\left[ i\(\phi-\phi_{\rm est}(y)\) \right]}\big\rangle_{y}\Big\rangle_{\phi}\bigg|^{-2}-1,
\eeq
Here $y$ is the data from which the estimate is made, and $\an{\cdots}_\bullet$ represents the ensemble average of the expression inside the angled brackets over $\bullet$.  This measure is miminized for the optimal estimate, 
\beq    \label{eqn:OptEst}
\phi_{\rm est}(y) = \arg  \big\langle{{\rm exp}\( i\phi\) }\big\rangle_{\phi|y},
\eeq
where $\an{\cdots}_{\bullet|y}$ means that there the average is conditioned (in a Bayesian way) on the data $y$. For this choice, which we will make henceforth, the Holevo deviation is equal to what has been dubbed the Holevo variance, 
\beq    \label{eqn:holvar}
V_{\rm H} = \Bigg(\Big\langle \big|\big\langle{{\rm exp}\( i\phi\) }\big\rangle_{\phi|y}\big| \Big\rangle_{y}\Bigg)^{-2}-1,
\eeq
This metric respects the cyclic property of the phase and sets the upper bound on the precision scaling in a sense that any other variance-like measure will scale at least as well as Holevo variance, whereas the reverse is not necessarily true~\cite{BerBre01}.

\subsection{Quantum phase estimation algorithm}

% ---------------------------------------------------------------------------------------
%	      	Heisenberg-limited optical phase estimation scheme
% ---------------------------------------------------------------------------------------

The HPEA is built upon the quantum phase estimation algorithm (QPEA) of Cleve \ea~\cite{CleMos98}; see \frf{fig:3photab}. At the core of QPEA lies an inverse quantum Fourier transformation (IQFT) that can be implemented with a scheme based on single-qubit inputs, single-qubit measurements and classical feed-forward on the reference phase $\theta$~\cite{GriNiu96}.  The Holevo variance of the QPEA scales theoretically as~\cite{BerWis09}
\beq \label{VarHol:QPEA}
V_{\rm H}^{\rm QPEA} = \frac{2}{N} + \frac{1}{N^2},
\eeq
which is above the SNL even for large $N$. The failure of the QPEA in this task can be understood by examining the probability distribution function $P(\phi_{\rm est})$ for the phase estimate, given by~\cite{BerWis09} 
         
\beq \label{PDF:QPEA}
P(\phi_{\rm est}) = \frac{1}{2\pi} \left|\sum_{n=0}^N {\cal C}_n e^{-i n (\phi_{\rm est}-\phi)} \right|^2,
\eeq
where ${\cal C}_n=\frac{1}{\sqrt{N+1}}$. This distribution profile (shown in \frf{fig:QPEA:dist}(a) for the case of $N=3$ and $N=7$) shows a sharp peak around $\phi_{\rm est} = \phi$, with width scaling as $1/N$ (as desired), but with relatively high wings. The envelope of the distribution for the wings falls off as the inverse square of the error in the estimate $(\phi_{\rm est}-\phi)^{-2}$, which gives rise to the leading order term in \erf{VarHol:QPEA}. Although QPEA can be improved by using a more complex adaptive measurement scheme~\cite{GioMac06,HigPry07}, even the generalized QPEA can only achieve  Heisenberg scaling in precision (with a constant overhead), but not the exact HL.

% -----------------------------------------------------------------------------------
%					Figure 
% -----------------------------------------------------------------------------------
\begin{figure}
	\includegraphics[scale=0.6]{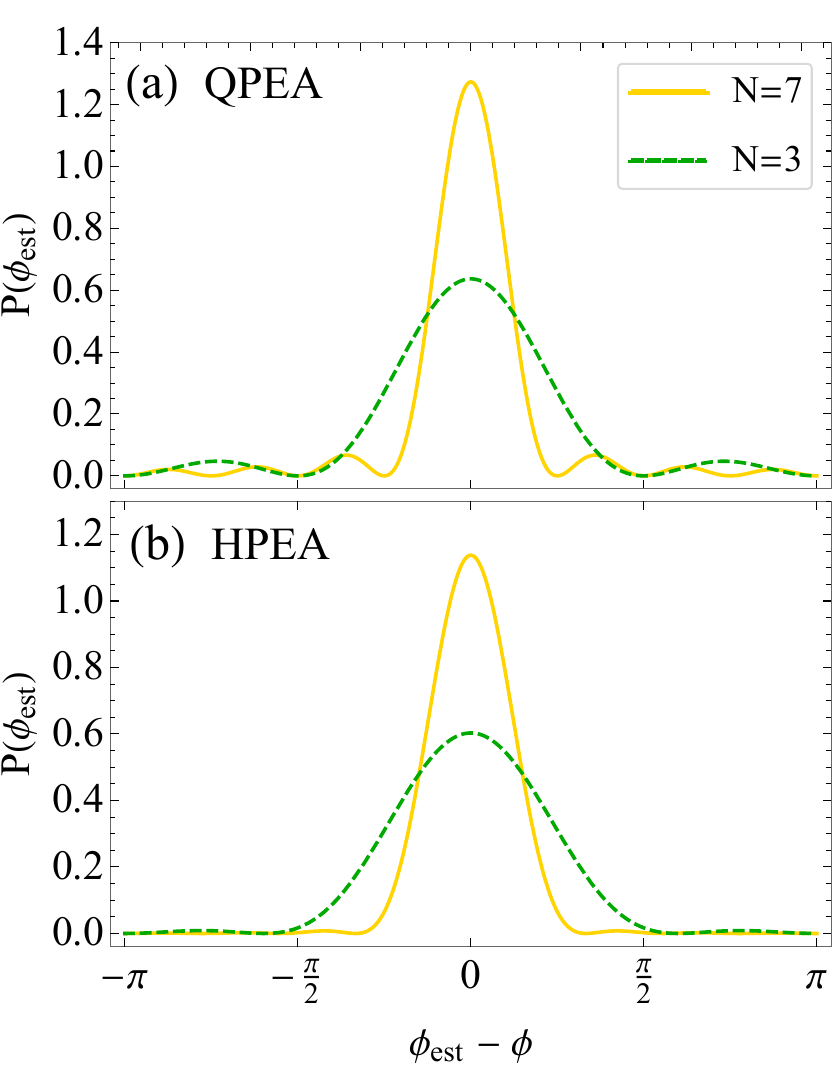}
	\caption{ \label{fig:QPEA:dist}(Color online) Probability distribution function for (a) the QPEA and (b) the HPEA  for two different numbers of quantum resources $N=3$ (dashed green), and $N=7$ (solid gold). It can be clearly seen that employing more resources results in a localized distribution function around $\phi_{\rm est}=\phi$. Optimizing the input state to the QPEA, the impact of rather high tails on phase estimation can be alleviated.}
\end{figure}

\subsection{Heisenberg-limited optical phase estimation scheme}   \label{sec:HPEA}

The key difference that allows HPEA to theoretically achieve the exact HL precision is the use of the optimal entangled probe state as the input to the QPEA. The quantum circuit for both the QPEA and HPEA for $N=7$ resources is shown in \frf{fig:3photab}(a). For the general case of $N=2^{K+1}-1$, a register of  $K+1=3$ qubits, is prepared in a product state for the QPEA, \frf{fig:3photab}(b), and in a particular entangled state for the HPEA, \frf{fig:3photab}(c). Then, $2^k$ CU gates are applied sequentially to each qubit followed, by the measurement in the $X$ basis, starting from the  qubit labelled by $k=K$. The result of the measurement classically controls reference rotation operations on the remaining qubits. The protocol proceeds downwards in $k$ until the qubit labelled by $k=0$ is measured. The $k$-th measurement yields the $k$-th bit in the binary expansion bit of $\phi/2\pi$. That is, the phase estimate is obtained according to the following relation
\beq  \label{phase:BinRep}
\phi_{\rm est}= 2\pi \times 0.\phi_0 \phi_1 \cdots \phi_K = 2\pi \,\mathlarger{\sum}_{k=0}^K \frac{\phi_k}{2^{(k+1)}}.
\eeq

For canonical (optimal) measurements the optimal state takes the following form~\cite{WisKil97} 
\beq   \label{S:opt:gen}
\ket{\Psi_{\rm opt}} \propto \sum_{n=0}^N\, \psi_n \ket{n},
\eeq  
where 
\beq
    \psi_n = {\rm sin} \[\frac{(n+1)\pi}{N+2}\],
\eeq 
$\ket{m}$ is the logical state of a register of qubits, and $m$ is a binary digit string of length $K+1$. The probability distribution function for estimating phase using this optimal state can be written in the same form as in \erf{PDF:QPEA} but with ${\cal C}_n = \psi_n/\sum_{n=0}^N |\psi_n|^2$. Note that the latter coefficients are in marked contrast to those of the QPEA, which are independent of $n$. 

The effect of input state optimization is to improve the phase estimation by reducing the high tails present in $P(\phi_{\rm est})$ for the QPEA. Figure \ref{fig:QPEA:dist}(b) illustrates this feature by plotting the probability distribution function for the phase estimate employing the optimal state, \erf{S:opt:gen}. 

Using this state the Holevo phase variance, \erf{eqn:holvar}, is minimized. That is, the interferometer attains its ultimate precision, the exact HL, which is expressed as~\cite{WisKil97} 
\beq  \label{Var:HL:exact}
V^{\rm HL} = {\rm tan}^2 \(\frac{\pi}{N+2}\).
\eeq

% -------------------------------------------------------------------------
%	     mode calculation for linear optical systems
% ------------------------------------------------------------------------- 
\subsection{Creating the optimal three-photon state} \label{sec:mmc}
For the $N=7$ case, using polarization degree of freedom to encode the qubits (\ie\ horizontal $\ket{{H}}\equiv\ket{0}$, vertical $\ket{{V}}\equiv\ket{1}$), the normalized optimal state \erf{S:opt:gen} can be written as
\beq   \label{s:opt:3phot}
\big|{\psi_{\rm opt}}\big\rangle = \sum_{j=0}^3 \alpha_j \ket{{\rm GHZ}_j},
\eeq
where $\ket{{\rm GHZ}_j}$ are Greenberger-Horne-Zeilinger (GHZ) type states \cite{GHZ93}
\begin{subequations}
	\bqa
	\ket{{\rm GHZ}_0}  &=&  \big(\ket{HHH} + \ket{VVV}\big) /\sqrt{2}, \label{ghz0} \\
	\ket{{\rm GHZ}_1}  &=&  \big(\ket{HHV} + \ket{VVH}\big) /\sqrt{2},  \label{ghz1} \\
	\ket{{\rm GHZ}_2}  &=&  \big(\ket{HVH} + \ket{VHV}\big) /\sqrt{2},  \label{ghz2} \\
	\ket{{\rm GHZ}_3}  &=&  \big(\ket{HVV} + \ket{VHH}\big) /\sqrt{2},  \label{ghz3}
	\eqa
\end{subequations}
and 
\begin{subequations} \label{S:opt3:cof}
	\bqa  
	\alpha_j &=& \sqrt{{2}/{\cal N}}\; {\rm sin} \big[ {(j+1) \pi}/{9}\big], \label{S:opt3:cofa}\\
	{\cal N} &=& 2 \sum_{j=0}^3 {\rm sin}^2 \big[ {(j+1) \pi}/{9}\big]. \label{S:opt3:cofb}
	\eqa
\end{subequations}
Here $\ket{HHH}$ denotes $\ket{H_{a}} \otimes \ket{H_{b}} \otimes \ket{H_{c}}$, and labels ``$a$'',``$b$'' and ``$c$'' correspond to the input qubits $k=0$, $k=1$ and $k=2$, respectively. The optimal state can be realized with the circuit depicted in \frf{fig:3photab}(c),
in which two CNOT gates are sequentially applied to the input state  $\ket{\psi_{\rm in}} = \ket{\psi_{a}} \ket{\psi_{bc}}$ where  $\ket{\psi_{a}} = \ket{+}$ and 
\beq    \label{psi:bc:gen}
\ket{\psi_{bc}} = \alpha_0 \ket{HH} + \alpha_1 \ket{HV} + \alpha_2 \ket{VH} + \alpha_3 \ket{VV},
\eeq
with $\alpha_i$ satisfying \erfs{S:opt3:cofa}{S:opt3:cofb}. Using $\hat U_{\rm CT}^{\rm CNOT}$ for the CNOT operation between the control (C) and target (T) qubit, the output state can be written
\beq  \label{psi:out:qb}
\ket{\psi_{\rm out}} = \hat U^{\rm CNOT}_{ac} \; \hat U^{\rm CNOT}_{ab} \,\ket{\psi_{\rm in}} \equiv \ket{\psi_{\rm opt}}.
\eeq

A concrete optical circuit to realize the state generation scheme is depicted in \frf{fig:circ:imper}. Here $\hat{a}_V\dots\hat{c}_H$ are the annihilation operators of the corresponding orthogonal polarization modes of photons ``$a$'',``$b$'' and ``$c$'', respectively and we use dual-rail encoding~\cite{RalPry10}, meaning that a single photon occupation of one of the orthogonal modes of the same photon corresponds to a logical  $\ket{0}$ or $\ket{1}$ state.

We use two types of probabilistic CNOT gates in our circuit. The $\hat U^{\rm CNOT}_{ab}$ operation is realized with a probabilistic non-universal CNOT (NCN) gate~\crf{PitFra02}. Here ``non-universality" means that it does not operate as a CNOT gate for a general two-qubit state but for a subset of bipartite state space. This gate is schematically shown in \frf{fig:circ:imper} inside the dashed-red box including four beam splitters with reflectivity $\eta_1 = \frac{1}{2}$ operating between the photons in modes ``$a$'' and ``$b$''. We note that the black and gray diamonds should be ignored throughout this section as they account for modeling imperfections which will be dealt with in \srf{sec:imperfect}. 

The task of the NCN is to entangle photons ${a}$ and ${b}$ by creating the state 
\beq \label{psi1:ab}
\ket{\psi_1} = \(\hat U^{\rm CNOT}_{ab}\otimes \hat I_{c}\) \,\ket{\psi_{\rm in}},
\eeq
which is the first part of $\ket{\psi_{\rm out}}$ in \erf{psi:out:qb}. Here $\hat I_{c}$ denotes an identity operation on mode labeled ``$c$''. With the input in modes ``$b$'' and ``$c$'' as \erf{psi:bc:gen} and $\ket{\psi_{a}} = \ket{H}$ the output of the NCN is 
\beq \label{psi:out:ab}
\bket{\psi_{\rm out}^{\rm NCN}} = \frac{1}{\sqrt{2}} (\ket{\psi_1} + \ket{\psi_{d}}), 
\eeq
where $\ket{\psi_{d}}$ represents a superposition of states with more than one photon in either of modes\footnote{{This depends on the value of the complex amplitudes $\alpha_j$ in \erf{psi:bc:gen}. For example, the mode labeled $\hat a_V$ can be in a two-photon state with probability $\smallfrac{1}{16}(|\alpha_2-\alpha_0|^2+|\alpha_3-\alpha_1|^2)$. }} and can be filtered out with an appropriate coincidence measurement. The result in \erf{psi:out:ab} shows that the state $\ket{\psi_1}$ is non-deterministically generated upon post-selection with probability success  $\wp_{\rm NCN} = \frac{1}{2}$.

The dashed-blue box in \frf{fig:circ:imper} shows a probabilistic universal CNOT (CN) gate~\cite{RalWhi02} between the two modes of the photon ``$c$'' (the target qubit) and the photon labeled ``$a$'' (the control qubit).  Photons in modes $a_{H}$ and $c_{V}$ non-classically interfere on the central (green) BS with reflectivity $\eta_2 = \frac{1}{3}$. There are two other such BSs, one of them located on the lower arm of the interferometer with the vacuum mode $v_{c}$ and the other one affecting only the control photon with the vacuum mode $v_{a}$. It was demonstrated in Refs.~\cite{BriBra03, LanWhi05} that  for a general two-photon state such as \erf{psi:bc:gen} this CNOT gate works with probability of success $\wp_{\rm CN}  = \frac{1}{9}$. 
By using the combination of NCN and CN the optimal state is non-deterministically obtained with the probability of success $\wp_{\rm opt} = \frac{1}{18}$. The post-selected successful operations correspond to the cases when each of the three outputs in \frf{fig:3photab}(c) contains at least one photon. Comparing to the circuit that uses two CN gates with overall probability of success of $\frac{1}{81}$, the advantage of using a NCN-NC circuit becomes clear. The detailed procedure for the calculation of the optimal state generation circuit  is described in \arf{appnA}.

% -----------------------------------------------------------------------------------
%					Figure
% -----------------------------------------------------------------------------------
\begin{figure}  
	\includegraphics[scale=0.42]{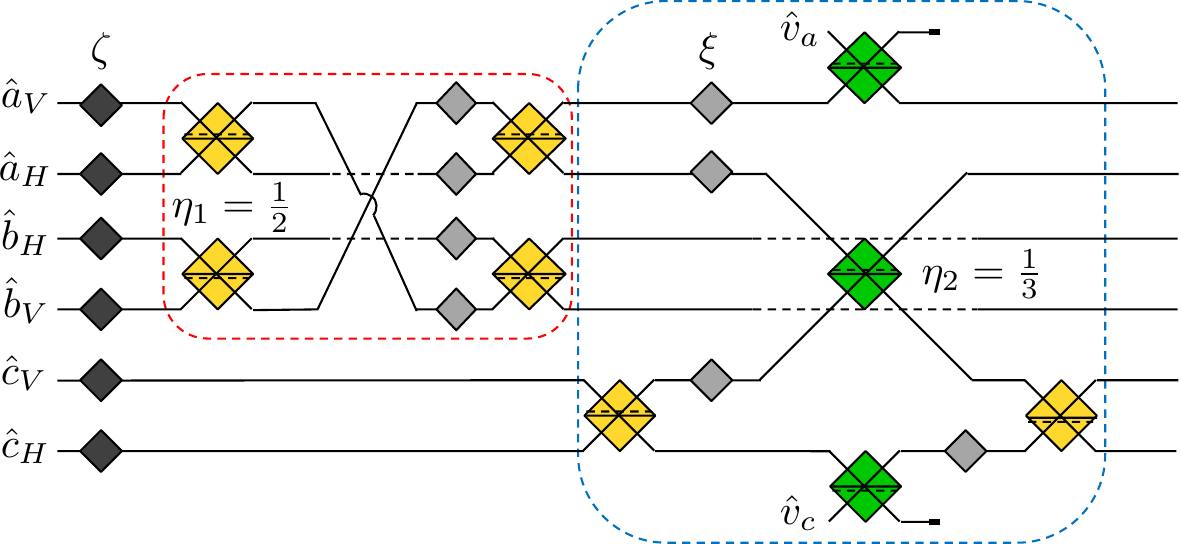} 
	\caption{\label{fig:circ:imper}(Color online) Conceptual circuit diagram for creating the optimal 3-photon state, \erf{s:opt:3phot}, consisting of two probabilistic CNOT gates. The dashed-red panel indicates a non-universal CNOT (NCN) gate operating between qubits labeled ``$a$'' and ``$b$'' where each is modeled by two polarization modes $H$ and $V$. The gold diamonds are beam splitters with reflectivity coefficient $\eta_1=\frac{1}{2}$; the dashed lines inside the beam splitters show that a photon reflected off that side acquires $\pi$ phase shift. Upon successful coincidence detection of photons this gates produces the state $\ket{\psi_1}$, \erf{psi1:ab}, with probability $\wp_{\rm NCN} = \frac{1}{2}$. The dashed-blue panel indicates a universal CNOT (CN) gate acting between qubits labeled ``$a$'' and ``$c$''. Each of these qubits has a vacuum port with an appropriate annihilation operator $\hat{v}_a$ and $\hat{v}_c$, respectively. The green diamonds are beam splitters with reflectivity $\eta_2 = \frac{1}{3}$. The gate successfully operates with probability $\wp_{\rm CN} = \frac{1}{9}$ due to post-selection. The black and gray diamonds illustrate beam splitters with reflectivity $\zeta$ and $\xi$,  respectively, for modeling total inefficiency in detecting photons and mode mismatch, respectively. Each of these BSs is treated in the same way explained in \srf{sec:imperfect}.} 
\end{figure}

% -----------------------------------------------------------------------------------
%					Figure
% -----------------------------------------------------------------------------------
\begin{figure}
\includegraphics[scale=0.58]{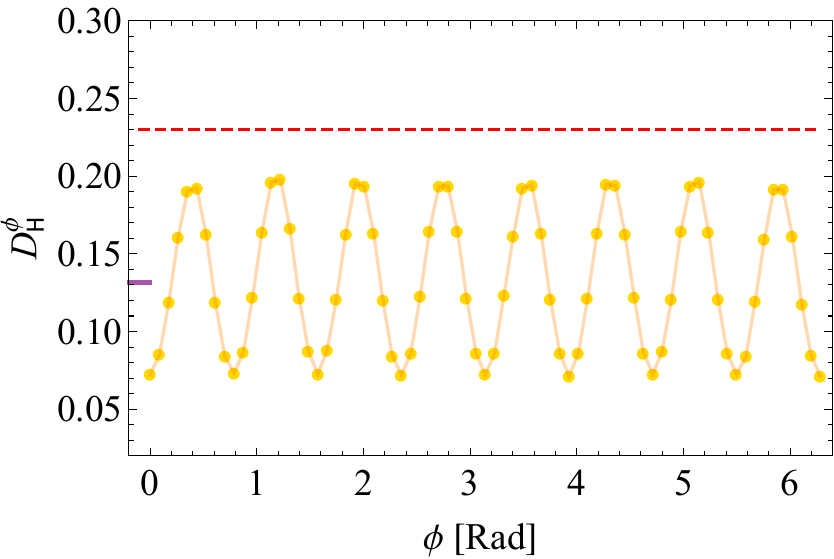}
\caption{ \label{fig:VH:all}(Color online) Variation of the phase-dependent deviation, \erf{eqn:HolVar:cond}, as a function of phase for the optimal  3-photon  state $\rho_{\rm opt}$, for the $K=2$ ($N=7$) HPEA (circular orange). For {\it ab-initio} phase estimation the Holevo deviation, \erf{eqn:holdev}, is used, an average which corresponds  to erasing any prior information about the phase. For measurements performed on the ideal state this is depicted by a purple horizontal line. The {\it ab-initio} SNL is shown by dashed red line. Note that since the probe state assumed here is perfect, the estimate from the HPEA would be optimal, \erf{eqn:OptEst}), so 
the Holevo deviation is equal to the Holevo variance, \erf{eqn:holvar}. }
\end{figure}

 % -----------------------------------------------------------------------------------
%					Figure 
% -----------------------------------------------------------------------------------
\begin{figure*}
\includegraphics[scale=0.68]{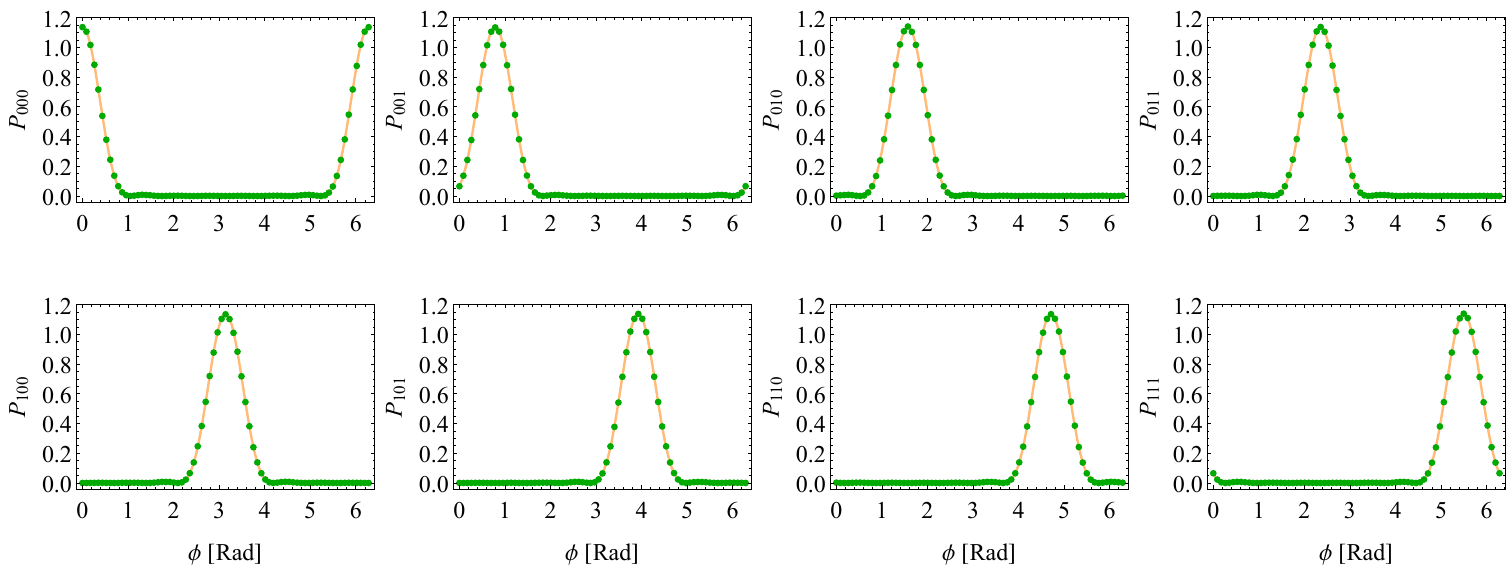}
\caption{ \label{pfixed:3phot}(Color online)  (Normalized) probability distribution, \erf{pdf:3phot}, of different measurement results as a function phase for the HPEA for $K=2$ ($N=7$). It is expected that when the true phase, $\phi$, is equal to one of the 8 possible binary digits sequences, $(\phi_0 \phi_1 \phi_2) \in \{000,001,\cdots,111\}$, the phase estimation algorithm's precision is at its best. The green dots are the results of numerical simulations, and the solid gold curves are obtained via \erf{PDF:QPEA} with the corresponding ${\cal C}_n$ for the HPEA. In all plots, $n_{\rm ens}=50\times10^3$.} 
\end{figure*}

% -------------------------------------------------------------------------
%	    		 Simulating the ideal protocol
% ------------------------------------------------------------------------- 
\subsection{Analysis of the HPEA} \label{sec:simtech}

At the end of the phase measurement protocol, each projective measurement yields one of the binary bits $\phi_k$ required for estimating the phase. The probability of getting a string $\phi_0 \phi_1 \cdots \phi_K$ of binary digits pattern, corresponding to one of the $2^{K+1}$ possible outcomes, is asymptotically ($n_{\rm ens} \rightarrow \infty$) equal to the number of times $n_{\phi_0 \phi_1 \cdots \phi_K}$ which that measurement result turns up divided by the size $n_{\rm ens}$ of the ensemble over which the Holevo deviation calculations are done:
\beq \label{pdf:3phot}
P_{\phi_0 \phi_1 \cdots \phi_K} = \frac{n_{\phi_0 \phi_1 \cdots \phi_K}}{n_{\rm ens}}. %n_{\phi_0 \phi_1 \phi_2}/n_{\rm ens}.
\eeq
To calculate the Holevo deviation a couple of more steps should be carried out. 
First, the ensemble average of $e^{i[\phi - \phi_{\rm est}(y)]}$ over the measurement results $y \in \{\phi_0 \phi_1 \phi_2\}$ needs to be worked out for calculating the deviation according to \erf{eqn:holdev}. That is, by choosing a phase in the interval $[0, 2\pi)$, and setting $\phi_{\rm est}(y)$ according to \erf{phase:BinRep}, one obtains the phase-dependent deviation\footnote{ Note that for the reconstructed state tomography $\phi_{\rm est}(y)$ is set slightly different than this ideal scenario, which will be explained in \srf{sec:exp}.}
\beq  \label{eqn:HolVar:cond}
D_{\rm H}^\phi = \Big|\big\langle{{\rm exp}\left[ i\(\phi-\phi_{\rm est}(y)\) \right]}\big\rangle_{y}\Big|^{-2}-1.
\eeq
\pu Then, \blk by erasing this initial phase information in the way defined in \erf{eqn:holdev} the phase-independent Holevo \pu deviation \blk $D_{\rm H}$ is determined. Figure~\ref{fig:VH:all} demonstrates simulation results for the optimal protocol using the state $\hat \rho_{\rm opt}$ given in \erf{s:opt:3phot}, for both of these quantities in orange dots and the purple horizontal line-segment cutting the left axis, respectively. The latter, up to an infinitesimal numerical error, is equal to the Heisenberg limit $V^{\rm HL} = 0.132474...$ obtained by letting $N=7$ in \erf{Var:HL:exact}. The procedure for stochastic simulations of the measurement circuit is described in \arf{appnB}. 

The shot-noise limit $V_{\rm SNL} = 0.232688...$ is depicted by the dashed-red line. In the SNL-limited measurement, uncorrelated photons are sent through the unknown phase and each is measured at its own measurement angle. To calculate the true SNL we minimize the variance  as function of these measurement angles. It is worth noting that if such minimization procedure is applied for the $N=3$ measurement, the corresponding SLN bound is $V_{\rm SNL} = 0.655845...$, instead of $V_{\rm SNL} = 0.777777...$, reported in \crf{DarPry18}, where the three photons were measured at different equidistant angles in $[0,2\pi)$. See \arf{appnSNL} for the details of calculations and measurement angles. 

As can be seen, the profile of $D_{\rm H}^\phi$ shows peaks and troughs around the HL with minima occurring at multiples of ${\pi}/{4}$. Each minimum equals one of the 8 estimated phase $\phi_{{\rm est}}$ that are possible from the 8 possible results $\phi_0 \phi_1 \phi_2$. 
This is also understood by analyzing the probability distribution for the above measurement output set; see \frf{pfixed:3phot}. For those phases for which $D_{\rm H}^\phi$ is minimum, the corresponding probability density function is maximum.  In contrast, for points in between minima, the phase-dependent deviation shows a less accurate estimation of the phase. Nevertheless, since we are interested in {\it ab-initio} phase measurement, the knowledge about phase can be removed by averaging over $\phi$ to obtain a precision value applicable to the entire range of $[0,2\pi)$.

The oscillatory behavior of $D_{\rm H}^\phi$ shown in \frf{fig:VH:all} depends on the number of resources employed. The higher $N$ means oscillations with a larger frequency and smaller amplitude.  For example, comparing the scenario here ($N=7$) with the two-photon $N = 3$ resources studied in \crf{DarPry18}, the number $N+1=2^{K+1}$ of oscillations doubled and the amplitude decreased by an order of magnitude.

% ============================================
%	     		 The effect of imperfections
% ============================================   
\section{The effect of imperfections}    \label{sec:imperfect}
 So far in our analysis of the Heisenberg-limited phase estimation algorithm everything took place in an ideal world. However, one should take into account practical considerations when it comes to realize such protocols in laboratories. This is typical in almost all physical experiments, and in particular, one of the important challenges in quantum metrology \cite{KacWal10, JacBan16, UlaLvo16, RocBar18}. Quantum-enhanced phase estimation schemes suffer from experimental imperfection in state preparation and detection. In quantum optics, these may be associated with optical mode mismatch (which leads to degraded nonclassical interference), the presence of multi-photon emission noise in imperfect single-photon sources, inefficient detectors, and the lack of photon-number resolving detectors. In what follows, we first present a model to address optical mode mismatch and imperfect detection and then separately consider the effect of multi-photon generation events on state preparation.  
 
% -------------------------------------------------------------------------
%	     	   Optical mode mismatch and loss
% ------------------------------------------------------------------------- 
\subsection{Optical mode mismatch} \label{sec:omm}
Central to the probabilistic CNOT gates considered in the previous section is the non-classical Hong-Ou-Mandel (HOM) interference phenomenon \cite{HOM87}. Two photons incident on a beam splitter perfectly interfere only if they cannot be distinguished from each other. This requires both of them to be in the same spatial, temporal, spectral and polarization modes. Any partial distinguishability of photons results in imperfect interference of quantum fields, which ultimately limits an experimenter's ability to prepare the desired optimized state.     

    % -----------------------------------------------------------------------------------
%					Figure (10)
% -----------------------------------------------------------------------------------
\begin{figure}
	\includegraphics[scale=0.47]{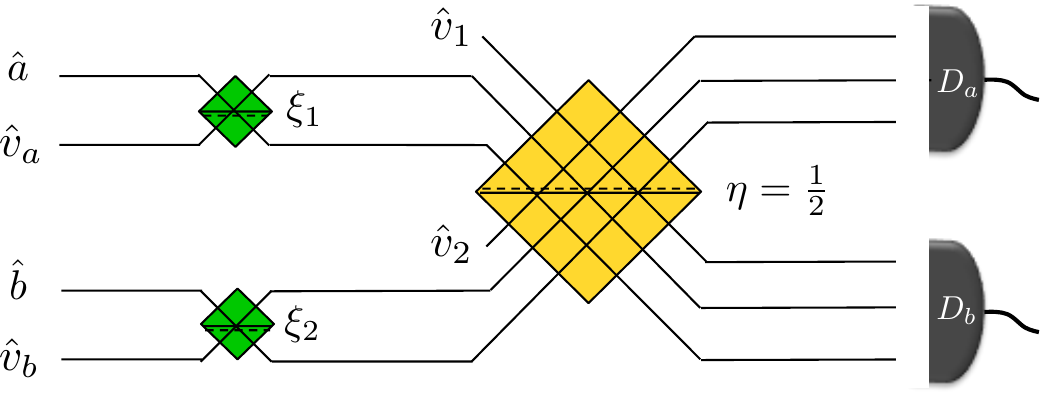}
	\caption{ \label{HOM:smm}(Color online)  Conceptual diagram of Hong-Ou-Mandel interferometer for modeling spatial optical mode matching. The imperfect overlap of modes is modeled by splitting the input beams (modes $\hat{a}$ and $\hat{b}$), via the gray BSs with reflectivities $\xi_j$, into two modes each such that only some portion of them interfere. $\hat{v}_j$ represent vacuum modes annihilation operators. In this case we show a 50:50 beamsplitter. } 
\end{figure}

One way to model mode mismatch in HOM interference on a beam splitter is illustrated in \frf{HOM:smm}. This model is employed in the state preparation shown in the circuit diagram, 
\frf{fig:circ:imper}. The two extra vacuum modes $v_{a}$ and $v_{b}$ are introduced for the main part of the incoming beams in modes $a$ and $b$, respectively. In addition, two other auxiliary vacuum modes $v_1$ and $v_2$ are included in the model for overlapping with $v_{a}$ and $v_{b}$, respectively. We characterize the degree to which the modes overlap by the BSs reflectivities $\xi_j (j=1,2)$ prior to the interference beam splitter. The ideal case obviously takes place when $\xi_j= 1$, for which the two modes perfectly overlap. This scheme is easiest understood in terms of spatial mode overlap, but is also applicable to modeling mismatch in any other single degree of freedom. 

In order to see how any mode mismatch leads to deteriorated quantum interference we calculate the probability $\wp_{\rm coin}$ of measuring a coincidence photon detection at detectors $D_{a}$ and $D_{b}$
\beq  \label{pcoin:smm}
\wp_{\rm coin} = \an{\hat{n}_{D_{a}} \hat{n}_{D_{b}}} = \frac{1}{2}\(1-\xi_1\xi_2\),
\eeq 
where $\hat n_{D_{a}}$ and $\hat n_{D_{b}}$ are the photon number operators 
for the bundle of three modes 
detected in each detector. The above expectation value is calculated using the input state $\ket{{\mathbb 0}}_{\hat{v}_1} \otimes \ket{{\mathbb 0}}_{\hat{v}_2} \otimes \ket{{\mathbb 0}}_{\hat{v}_3} \otimes \ket{{\mathbb 0}}_{\hat{v}_4} \otimes \ket{{\mathbb 1}}_{\hat{a}} \otimes \ket{{\mathbb 1}}_{\hat{b}}.$ Here $\ket{{\mathbb 0}}$ and $\ket{{\mathbb 1}}$ denote the vacuum and single photon state, respectively.That is, the two modes indicated by $\hat a$ and $\hat b$ are in a single photon state and the the rest are in the vacuum state. 
Ideally, a coincidence detection happens between modes ${a}$ and ${b}$ whereas for the scenario sketched here 9 coincidence detection possibilities occur. The algebra leading to \erf{pcoin:smm} is straightforward and can be found in \arf{appnC}. Note that for the perfect mode matching $\xi_1=\xi_2=1$ the coincidence probability is zero as expected in an ideal HOM interference phenomenon. This probability also acquires its classical value of $\wp_{\rm max}=\frac{1}{2}$ for the case of no overlap of the modes when either of the BSs are totally transmitting, $\xi_j=0$.  

A good measure that can capture the effect of mode mismatch is the quality of interference fringes or non-classical interference of photons known as dip visibility,
\beq    \label{HOM-dip:vis}
\nu = [\wp_{\rm max}-\wp_{\rm coin}]/\wp_{\rm max} = \xi_1 \xi_2.
\eeq
From this relation it can be easily seen that for an ideal overlapping of modes where both auxiliary BSs are totally reflecting the visibility is $1$ and it becomes zero for $\xi_j=0$ where the two modes completely mismatch. The worse the mode matching, the greater the deviation of dip visibility from its ideal value of 1.

% -----------------------------------------------------------------------------------
%					Figure (11)
% -----------------------------------------------------------------------------------
\begin{figure} 
	\center
	\includegraphics[scale=0.9]{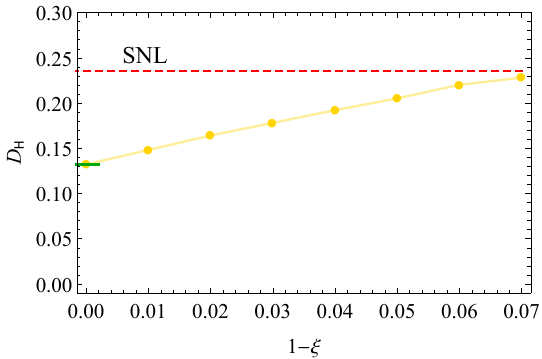}
	\caption{ \label{HolVar:smm}(Color online)  Impact of optical mode mismatch on the overall performance of the HPEA. The result of numerical simulations of the Holevo deviation, \erf{eqn:holdev}, is illustrated using golden solid points (solid golden line is for guiding eyes). The green dashed line depicts the SNL. Heralding efficiency is set at 13\% for all simulations and each point is obtained using $10^4$ runs. } 
\end{figure}

Now this model can be incorporated into the mode calculation analysis for the optimal state preparation. An implication of this is that the generated state will no longer be the same as $\rho_{\rm opt}$ and as a consequence the protocol's performance in estimating phase drops. In order to see this effect we place some beam splitters with reflectivity $\xi$ at relevant beam paths, shown by gray diamonds in \frf{fig:circ:imper}. This means some auxiliary modes are introduced in the same way discussed above. Note that the inserted BSs do not have to be identical. Nethertheless, assuming alike BSs is a good approximation of a real experimental scenario. Another type of auxiliary BSs with reflectance coefficient $\zeta$, illustrated in  \frf{fig:circ:imper} as black diamonds is used to model non-unit efficiency of the photon detectors. One can introduce these latter BSs in the very beginning of the circuit, due to the linearity of optical elements which allow us to shift them through all the way from the detection stage.

Incorporating all these auxiliary beam splitters and modes into the state preparation circuit allows for conducting numerical simulations\footnote{ Here instead of sweeping $\phi$ in some increments in the entire interval $[0,2\pi)$ and working out the ensemble averages over $y$ and $\phi$ as in \erf{eqn:holdev}, the true phase is chosen randomly in that interval for each execution of the circuit. Then the resulting string $y\equiv\phi_0 \phi_1 \phi_2$ determines $\phi_{\rm est}(y)$ according to \erf{phase:BinRep}. This leads to calculating one instance of the exponential term in \erf{eqn:holdev}. An ensemble average of sufficiently many instances yields the expression inside the modulus squared which is used to compute the Holevo deviation.} to determine the Holevo deviation for different amount of mode mismatch. This is depicted in \frf{HolVar:smm} where the dashed green line shows the SNL and golden data points are the results of numerical simulations. Supposing there are no other experimental imperfections, the sub-shot-noise precision would not be observed for mode mismatch above about $7\%$. In an ideal condition $\xi=1$ the protocol performs at the exact Heisenberg limit, as expected.

% -------------------------------------------------------------------------
%	      Higher order terms in the SPDC process
% ------------------------------------------------------------------------- 
\subsection{Higher order terms in the SPDC process}
The output state of a SPDC process in the photon-number basis can be expressed as the product of the down-converted photons state $\ket{\psi}_{\rm SPDC}$ and pump photon state where \cite{MigBie13}
\beq \label{state:spdc}
\ket{\psi}_{\rm SPDC} \approx \ket{{\mathbb 0} {\mathbb 0}} + \epsilon \ket{{\mathbb 1} {\mathbb 1}} + \frac{\epsilon^2}{2} \ket{{\mathbb 2} {\mathbb 2}} + O(\epsilon^3).
\eeq
Here $\ket{n_{\rm s} n_{\rm i}}$ represents the photon-number basis with the signal and idler photons $n_{\rm s}$ and $n_{\rm i}$, respectively. The parameter $\epsilon$ is an overall efficiency related to the pump power, the nonlinear constant, and SPDC crystal thickness. \erf{state:spdc} shows the non-deterministic nature of these conventional photon sources where $\epsilon$ determines the rate of producing single photons. Multiple-photon generation events usually contaminate the quantum state and, if the photon detector does not possess the photon number resolution capabilities,  we cannot distinguish between single- and few-photon detection.

% ----------------------------------------------------------------------------------
%					Figure (7_12)
% -----------------------------------------------------------------------------------
\begin{figure}
	\center
	\captionsetup[subfigure]{labelformat=empty}
	\subfloat[]{\includegraphics[scale=0.86]{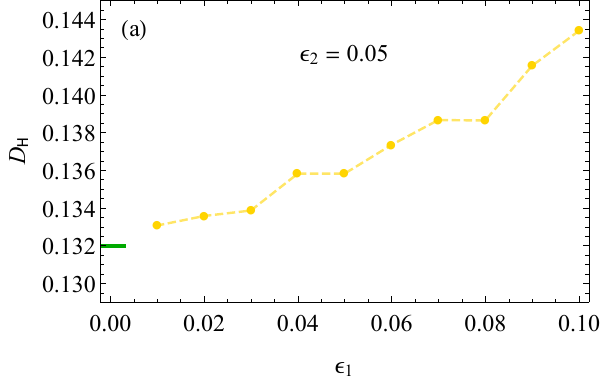} \label{VH:hots:e1} } \\[-0.4cm]
	\subfloat[]{\includegraphics[scale=0.86]{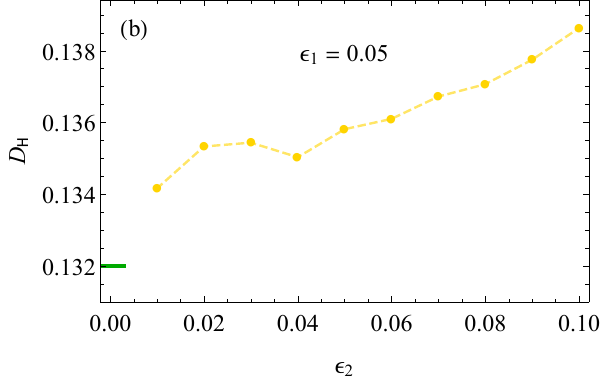} \label{VH:hots:e2}} 
	\caption{\label{VH:hots:e12}  The HPEA performance in the presence of higher-order terms in the SPDC process. The Holevo deviation, \erf{eqn:holdev}, is plotted by varying the overall efficiency of (a) the first SPDC source, while $\epsilon_2 = 0.05$, and (b) swapping the role of $\epsilon_1$ and $\epsilon_2$. The heralding efficiency is fixed at 13\% for both plots. Each data point was obtained using $50\times10^3$ simulation runs.}
\end{figure}

To create three photons required for the experiment two type-I SPDC sources are employed. The first one is supposed to supply horizontally polarized single photons for mode ``$a$" which are heralded by their partners in the trigger mode ``$t$". The output state of this source can be written as
 \beq   \label{psi:at}
\ket{\bar\psi_{at}} \approx \Big(1 + \epsilon_1 \hat{a}_H\dg \hat{t}\dg + \frac{\epsilon_1^2}{2} \hat{a}_H^{\dagger\,2} \hat{t}^{\dagger\,2} + \frac{\epsilon_1^3}{6} \hat{a}_H^{\dagger\,3} \hat{t}^{\dagger\,3}\Big) \ket{{\mathbb 0} {\mathbb 0}},
\eeq 
where the overhead bar indicates that the resultant state contains multiphoton terms. Here, the terms higher than third order in $\epsilon_1$ are discarded.

%---------------------------------------------------
%					Figure 
% --------------------------------------------------
\begin{figure*}
	\includegraphics[scale=1.0]{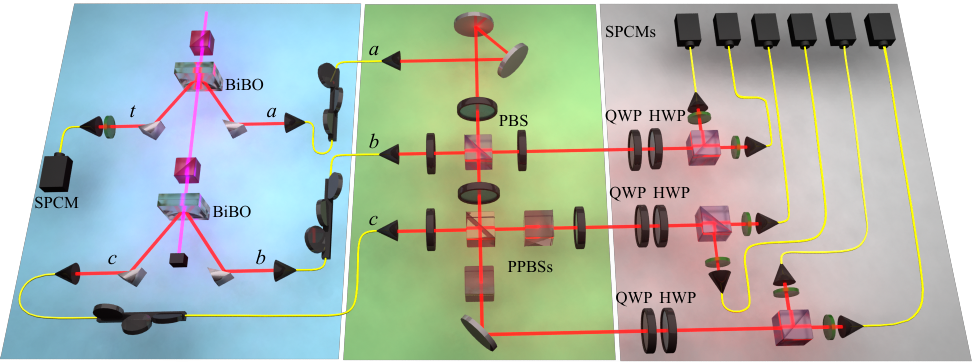}
	\caption{\label{fig:setup}  Experimental setup arrangement. (Blue region) Single and entangled photons at $820{\rm nm}$ are generated via two type-I SPDC sources. The SPDC crystals are pumped by pulsed UV light produced through second harmonic generation process. Photons are guided using single-mode fibers into the entangling gate to create the optimal state. (Green region) The desired probed state is post-selectively generated by realizing two entangling gates: the probabilistic non-universal CNOT gate acting between modes ``$a$" and ``$b$" composed of four HWPs and one PBS (equivalent to the red dashed box in \frf{fig:circ:imper}), and non-deterministic universal CNOT gate operating between modes ``$a$" and ``$c$" made up of three PPBSs where the central one (non-flipped) coherently combines the control and target photons and two HWPs set at $22.5^\circ$ with respect to the optical axis (corresponding to the blue dashed box in \frf{fig:circ:imper}). (Gray region) quantum state reconstruction tomography stage. Photons are directed to polarization analysis units consisting of a QWP, HWP and PBS followed by a 2nm spectral filter and SPCM. }
\end{figure*}

A second type-I SPDC source (composed of two sandwiched BiBO crystals with perpendicular optical axes)  with overall efficiency  $\epsilon_2$  provides photons for modes ``$b$" and ``$c$" to realize the state $\ket{\psi_{bc}}$ given by \erf{psi:bc:gen}. Preparing the pump photon in a linearly polarized state $\beta\ket{H}+\gamma\ket{V}$ where $\abs{\beta}^2 + \abs{\gamma}^2 =1$, the state of down-converted photons can be described (up to second order in $\epsilon_2$) by \cite{KwiEbe99}
 \bqa  \label{psi:spdc2} 
 \ket{\bar\Xi} &\approx&  \bigg[1+ \epsilon_2\(\gamma\, \hat{b}_H\dg \hat{c}_H\dg + \beta\, \hat{b}_V\dg \hat{c}_V\dg\) + \nonumber \\
 &&\hspace{10pt} \frac{ \epsilon_2^2}{2} \(\gamma \,\hat{b}_H\dg \hat{c}_H\dg + \beta\, \hat{b}_V\dg \hat{c}_V\dg\)^2 \bigg] \ket{{\mathbb 0} {\mathbb 0}}.
 \eqa
The multiphoton state $\ket{\bar\Xi}$ needs to be converted into $\ket{\bar\psi_{bc}}$, for which the single-excitation terms form the desired state $\ket{\psi_{bc}}$ of \erf{psi:bc:gen}. This can be achieved by subjecting $\ket{\bar\Xi}$ to further unitary evolution by means of two linear optical elements, that each implement a tuneable beam splitter operation in polarization basis on a specific mode ``$m$''
 \beq
  \hat{\mathfrak U}(\vartheta_{m}) = {\rm exp}\[-i\vartheta_{m}\(\hat{m}_H\dg \hat{m}_V + \hat{m}_H \hat{m}_V\dg\)\],
 \eeq
 with ${m} = \{{b},{c}\}$.
 The values for $\vartheta_b$ and $\vartheta_c$ are chosen so that one obtains
 \beq \label{psi:bc:hot}
 \ket{\bar\psi_{bc}} = \hat{\mathfrak U}(\vartheta_{b})\, \hat{\mathfrak U}(\vartheta_{c})\, \ket{\bar\Xi}.
 \eeq
Now, using \erfa{psi:at}{psi:bc:hot} the input state prior to optimal state preparation gate is 
 \beq
 \ket{\bar\psi_{\rm in}} = \ket{\bar\psi_{at}} \ket{{\bar\psi}_{bc}},
 \eeq 
 where only 3-photon amplitude terms $\epsilon_1 \epsilon_2, \epsilon_1 \epsilon_2^2, \epsilon_2 \epsilon_1^2, \epsilon_1^3 $ are used for post-selecting 3-click coincidence detection and higher order amplitudes are neglected.  
 
Applying this approach and assuming that optical modes perfectly overlap, numerical simulations of the phase measurement protocol can be accomplished through the same procedure as in the previous section. Before doing so, we need to determine two parameters $\epsilon_1$ and $\epsilon_2$ for the calculations. The overall efficiency of a pulsed SPDC source is related to its coincidence count rate ${\mathfrak C}$ via \cite{AguKov14} 
 \beq
\epsilon = \sqrt{ \frac{ {\mathfrak C}}{{\mathfrak R}\, \lambda_{\rm i} \lambda_{\rm s}}},
 \eeq
where ${\mathfrak R}$ is the pulsed laser repetition rate, $\lambda_{\rm i}$ and $\lambda_{\rm s}$ are the heralding efficiency of the idler and signal modes. For the experiments conducted in this work we found the overall efficiency approximately satisfying $\epsilon \in [0.05, 0.1]$. In particular, for the setting of $100 {\rm mW}$ pump power, ${\mathfrak R}=80 {\rm MHz}$, $\lambda_{\rm i}\approx\lambda_{\rm s}= 13\%$ and ${\mathfrak C}\approx 5200$ we obtain $\epsilon \approx 0.06$. The result of computational modeling for the Holevo deviation $D_{\rm H}$ for varying $\epsilon_1$ in the above interval while keeping $\epsilon_2=0.05$ is illustrated in \fsrf{VH:hots:e12}{VH:hots:e1}. The detection efficiency is set $13\%$ as before. In \fsrf{VH:hots:e12}{VH:hots:e2} the role of the two overall efficiencies are swapped. 

Drawing a comparison between these plots and \frf{HolVar:smm} it is obvious that optical mode mismatch would be expected to have more impact on the performance of the phase measurement scheme. As mentioned before, all of these experimental imperfections may be present at the same time. We avoided including them all together due to computational costs. However, in order to get a proper account of real experimental situation we will evaluate the protocol performance using the experimentally generated state,  reconstructed via quantum state tomography.

% -------------------------------------------------------------------------
%	         		   Experiment
% -------------------------------------------------------------------------
\section{Experimental realization of the probe state} \label{sec:exp}
In this section we present the experimental implementation of the optimal state creation. The experimental configuration is schematically shown in \frf{fig:setup} consisting of three sections: the single-photon sources (blue panel), entangling gate for preparing the optimal state (green panel), and quantum state tomography stages (gray panel). Two cascaded type-I SPDC sources (described elsewhere~\cite{DarPry18})) are employed to supply photonic qubits encoded in the polarization degree of freedom. A single photon in mode ``$a$" is heralded by its partner in the trigger detector produced by the first source. The second one generates a pair of entangled photons which are directed via fiber coupling towards modes ``$b$" and ``$c$" in the state preparation gate.  

The NCN gate (red-dashed box in the circuit diagram in \frf{fig:circ:imper}) is realized  with a polarization beam splitter (PBS) and four half waveplates (HWPs). The CN gate  (blue-dashed box in the circuit diagram in \frf{fig:circ:imper}) is made up of three partially-polarized beam splitters (PPBSs) and two HWPs. The central PPBS has reflectivity $\eta_V = \frac{2}{3}$ and $\eta_H = 0$ for the vertically and horizontally polarized light, repectively. The other two PPBSs were flipped by 90 degrees around the propagation direction of photons such that $\eta_V = 0$ and $\eta_H = \frac{2}{3}$; see \frf{fig:setup}. The two HWPs serve as $50:50$ beam splitters of the polarization interferometer.

% ----------------------------------------------------------------------------------
%					Figure 
% -----------------------------------------------------------------------------------
\begin{figure}[b]
	\includegraphics[scale=0.55]{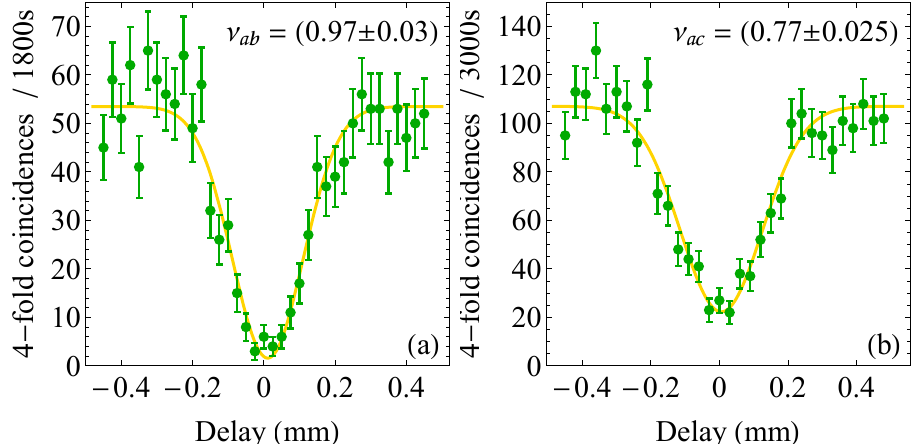}
	\caption{\label{HOMs}  Experimental HOM interference of photons produced via two independent SPDC sources. (a)  Photons in modes ``$a$" and ``$b$" non-classically interfere in the PBS; see \frf{fig:setup}. The observed visibility is $\nu = 0.97\pm0.03$, and the dip width is $229\pm19\,{\rm \mu m}$. (b) Interference between photons in modes $a$ and $c$ in the central PPBS. The observed dip visibility is  $\nu = 0.77\pm0.025$, and the dip width is $228\pm14\,{\rm \mu m}$. See text for the details.}
\end{figure}

We saw in \srf{sec:omm} that high-quality quantum interference of photons is crucial for creating a state close to the optimal state, \erf{s:opt:3phot}. This phenomenon takes place in our experiments between polarized photons in modes ``$a$" and ``$b$" incident on the PBS (in the NCN gate) and between photons in modes ``$a$" and ``$c$" impinging on the central PPBS (in the CN gate). Note that even though HOM interference occurs between two photons, the presence of these two is heralded by the other photon pair each from an independent SPDC source. This means 4-fold coincidence counting which leads to increasing the data collection time during which the setup should remain stable. The maximum interference visibility that can be obtained for a PBS is $1$ and that of a PPBS with $\eta_V = \frac{2}{3}$ is $0.8$~\cite{DarPry18}. We have observed $\nu_{ab} = 0.97\pm0.03$ and $\nu_{ac} = 0.79\pm0.025$ visibility HOM interference, respectively; see \frf{HOMs}.

\begin{figure}
\captionsetup[subfigure]{labelformat=empty}
\subfloat[]{\includegraphics[scale=0.5]{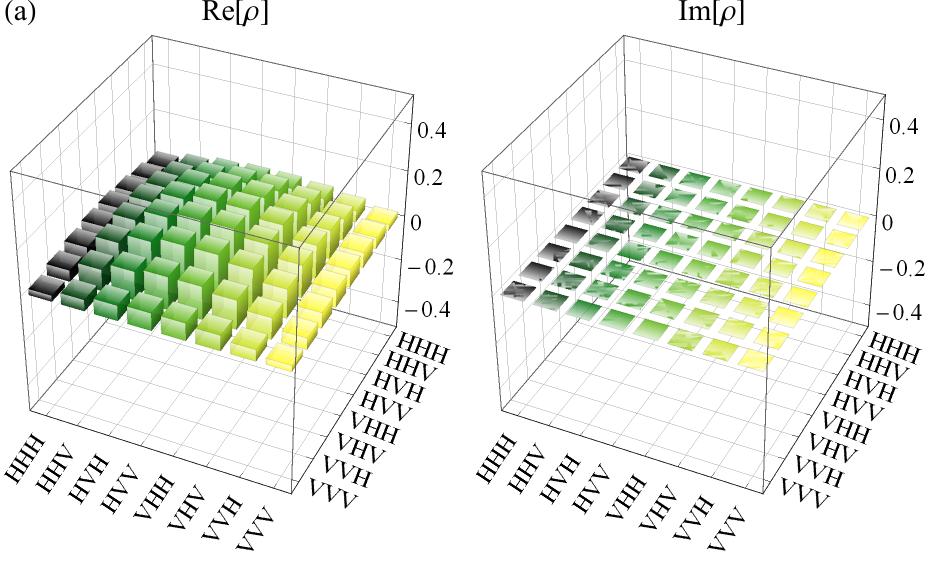} \label{tomo_theo} } \\[-0.5cm]
\subfloat[]{\includegraphics[scale=0.5]{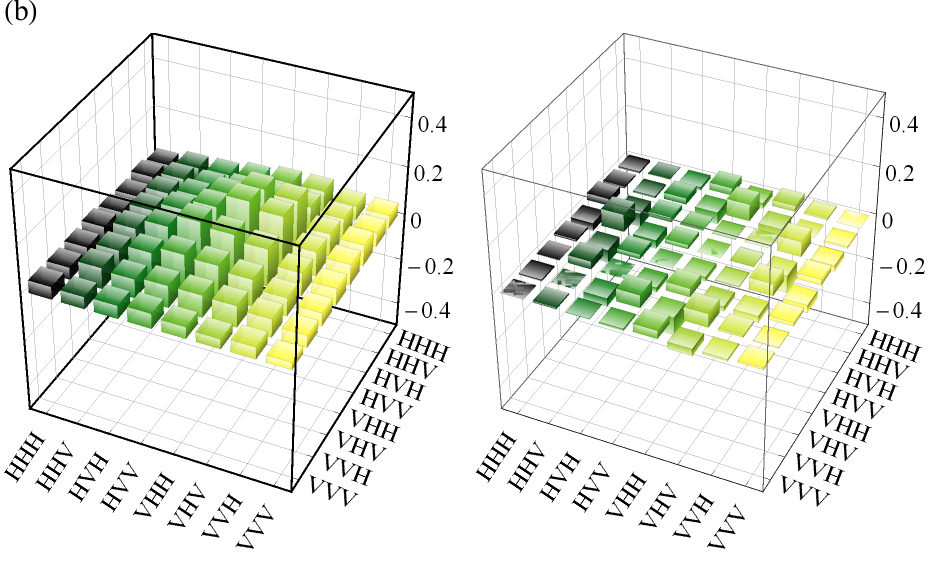} \label{tomo_exp}} 
\caption{\label{fig:tomo}(Color online) (a) Real (left) and imaginary (right) parts of the state matrix $\rho_{\rm opt}$ reconstructed from polarization state tomography, and (b) the optimal state $\rho_{\rm exp} = \op{\psi_{\rm exp}}{\psi_{\rm exp}}$, \erf{s:opt:3phot}. The fidelity of the experimental state with respect to the optimal state is $F = 0.810\pm0.014$, and the purity is equal to ${\cal P} = 0.75\pm0.02$, calculated from approximately 4200 fourfold coincidence photodetection.}
\end{figure}

Finally, to characterize the state, three-qubit polarization quantum state tomography was performed~\cite{WhiLang07}. Figure \ref{fig:tomo}\subref{tomo_theo} demonstrates the real and imaginary parts of $\rho_{\rm opt} = \op{\psi_{\rm opt}}{\psi_{\rm opt}}$ calculated from \erf{s:opt:3phot}. The result of reconstructed quantum state tomography $\rho_{\rm exp}$ using a maximum likelihood estimation technique is shown in \fsrf{fig:tomo}{tomo_exp}. The state fidelity \cite{Joz94} with respect to $\rho_{\rm opt}$ was measured to be $F =  0.810\pm0.014$, and the state purity ${\cal P} = {\rm Tr}\big[\rho_{\rm exp}^2\big] = 0.75\pm0.02$. Uncertainties are estimated using a Monte Carlo numerical simulation, sampled from a Poisson distribution of photon counts. The measurements were taken at the low pump power setting, to ensure small, $\epsilon\approx0.06$, probability amplitude of generating more than one photon pair from the same source. Together with a heralding efficiency of $\approx 0.13$, characteristic for non-collinear SPDC sources, this resulted in a low count rate, typically on the order of few four-fold coincidences per minute. The overall state quality is comparable (or better) with the states obtained by optical circuits that involve two CNOT gates and similar state generation technology ({\it e.g.}~\crf{vitelli13}).

% -----------------------------------------------------------------------------------
%					Figure (15)
% -----------------------------------------------------------------------------------

%-----------------------------
%					Figure
% ---------------------------------------------
\begin{figure}
\includegraphics[scale=0.58]{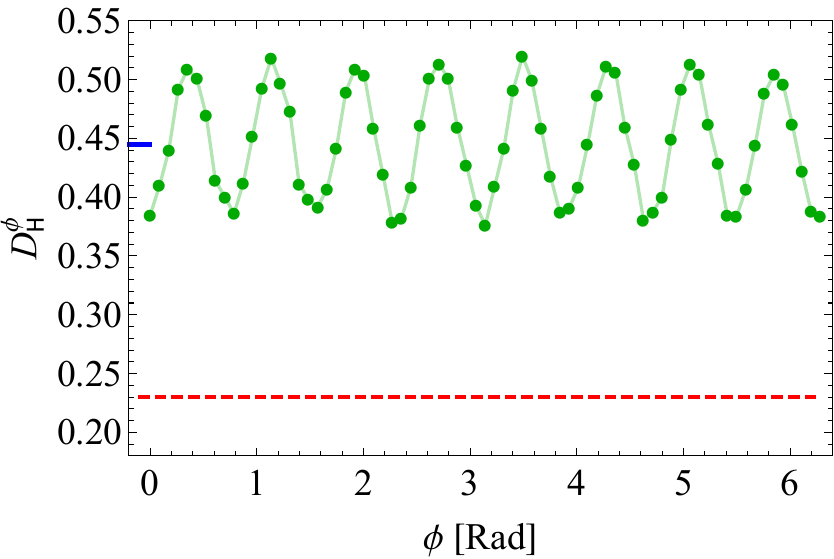}
\caption{ \label{fig:VH:exp}(Color online) Profile of the phase-dependent Holevo deviation, \erf{eqn:HolVar:cond}, as a function of phase for the experimental state $\rho_{\rm exp}$ (rectangular green points). The Holevo deviation, \erf{eqn:holvar}, is depicted by a blue horizontal line. The dashed red line illustrates the SNL as in \frf{fig:VH:all}.}
\end{figure}

We can test the expected performance of the obtained experimental state $\rho_{\rm exp}$ in the {\it ab-initio} phase measurement protocol. To this end, numerical simulations of the algorithm was executed by replacing for the optimal state in the scheme and following the same recipe discussed in \srf{sec:simtech} and \arf{appnB}\footnote{ There is an exception here. The procedure outlined in \srf{sec:simtech} has been formalised for the optimal state. However, as it will be seen in \frf{pfixed:3phot:exp}, the reconstructed state tomography lead to probability distributions that suggest $\phi_{\rm est}(y)$ are slightly shifted from those given in \erf{phase:BinRep}. The difference between any of the latter and corresponding expression ${\rm arg}(\int P_{\phi_0\phi_1\phi_2} e^{i\phi} d\phi)$, where $P_{\phi_0\phi_1\phi_2}$ is the normalized probability distribution, results the small shift. }. The results are illustrated in \frf{fig:VH:exp}. The phase-dependent deviation $D_{\rm H}^\phi$ reveals the oscillatory behavior (green square markers) similar to the ideal state case, but unfortunately well above the SNL. After averaging over the true phase, the Holevo deviation is determined to be $D_{\rm H} = 0.445$ (blue horizontal line-segment), compared to the SNL of 0.232688.

These experimental results contrast with the predictions of our simulation results of \frf{HolVar:smm} and \frf{VH:hots:e12}, that show that for the achieved levels of mode overlap and high-order photon number noise, $\xi\approx 0.98$ and $\epsilon \approx 0.06$, the generated state should perform better than the SNL. The corresponding probability distribution of different measurement outcomes, \frf{pfixed:3phot:exp}, simulated 
using $\rho_{\rm exp}$ as input, show a minor deviation of the function profile from the ideal bell-shaped curve and lower than expected maximum probability density. The small bump in the dotted curves hints there should be a slight shift in the locus at which maximum probability occurs in comparison to those of the perfect case. As a result, the estimate $\phi_{\rm est}(y)$ should account for this slight difference with respect to values obtained from \erf{phase:BinRep}.

\begin{figure*} 
\includegraphics[scale=0.68]{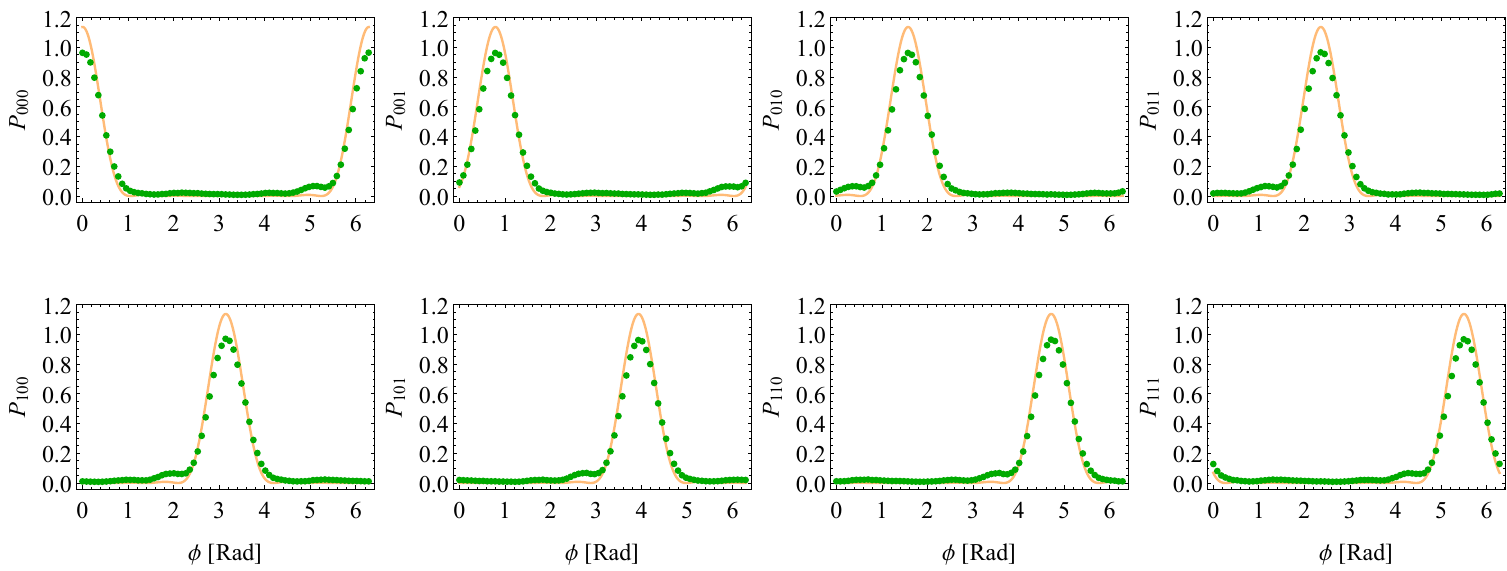}
\caption{ \label{pfixed:3phot:exp}(Color online) (Normalized) probability distribution as in \frf{pfixed:3phot}, but the green dots are simulation results using the reconstructed state $\rho_{\rm exp}$. The solid gold curves are obtained via \erf{PDF:QPEA} with the corresponding ${\cal C}_n$ for the HPEA. In all plots, $n_{\rm ens}=50\times10^3$.} 
\end{figure*}

With this insight, we identify several causes for the discrepancy between the observed results and our theoretical prediction. First, our simulations were performed under the assumption that only a single source of noise was present and did not take into account combined effect of both mode overlap and high-photon number noise on the probe state quality. Second, the low levels of noise were achieved at the expense of extremely low count rate due to low pump power and high loss from spectral filtering, leading to long measurement times. Although HOM interference demonstrated high mode overlap, this measurement was taken over a relatively (under 20 hours) short time period. Three-qubit quantum state tomography lasted significantly longer (more than a week), leading to an inevitable HOM dip position shift, and degradation of quantum interference and overall optical setup alignment, impacting the probe state beyond what was predicted by the initial numerical simulation. Finally, despite generating a probe state of reasonably high-quality in terms of noise, a systematic deviation from the ideal form of \erf{s:opt:3phot} would also degrade its performance in phase estimation by distorting the measurement outcome probability density functions from the ideal shape, as shown in \frf{pfixed:3phot:exp}. Such deviations are generally addressed through a calibration routine, consisting of iterations of measurements and setup adjustments. This approach was impractical in our case due to the very low count rates. One way to overcome these challenges would be through increasing the count rates and time and frequency mode overlap by implementing novel efficient single-photon sources \cite{WesPry16,SenWhi17,SluPry19}. The same result cannot be simply achieved by boosting up the pump power as it would increase the probability of producing unwanted multi-photon pairs.

% -------------------------------------------------------------------------
%				Results and discussion
% -------------------------------------------------------------------------
\section{Conclusion}   \label{sec:results}
In summary, we studied a Heisenberg-limited interferometric phase estimation algorithm in the presence of experimental imperfections. Although, in principle, the protocol can exploit any physical qubit, our attention was concentrated to experimental realization using optical photons. Our optical scheme for preparation of the optimal three-photon $N=7$ state included two probabilistic CNOT gates for which quantum interference of photons plays a pivotal role.  

\ We numerically analyzed the effects of the two major  expected experimental imperfections of the photonic setup:  
mode mismatch of the interfering photons and multi-photon generation events in the photon sources. We found the impact of the former  
to be more severe than the latter.  Nevertheless, we predicted that sub-SNL phase estimation should be possible in presence of experimentally achievable levels of either type of imperfection, individually.

 However, when we attempted to experimentally realize %  realization of 
the optimal probe state, and characterized it by tomography, a numerical simulation of its use in phase estimation, predicted a performance for phase estimation worse than the shot-noise limit.  This discrepancy can perhaps be attributed to the fact that the necessary low levels of noise were achieved at the expense of very low count rates, which translate into long measurement duration making it impractical to both keeping high-quality of HOM interference for the entire duration of the tomography. This problem of stability would also likely affect the calibration of the state for the phase estimation protocol.

Nascent efficient photon sources, detection technology and advances in integrated optics should be able to tackle this problem~\cite{SluPry19}. Upon overcoming  this obstacle, future works can be directed towards implementing the full phase measurement circuit, which requires fast feedback control operations.

% -------------------------------------------------------------------------
%				acknowledgement
% -------------------------------------------------------------------------
\section*{Acknowledgement}
The authors are indebted to Jian Li, formerly of the University of Science and Technology of China, for work on the state preparation protocol of Sec.~\ref{sec:mmc}.  We also thank Raj B. Patel and Allen Boston for assistance with data acquisition code and Dominic W. Berry for useful discussions. This research was supported by the Australian Research Council Centre of Excellence Grants No. CE110001027 and CE170100012. S.D. acknowledges financial support through an Australian Government Research Training Program Scholarship and thanks the University of Western Australia.

% ---------------------------------------------------
%			Appendix A
% ---------------------------------------------------
\appendix
\numberwithin{equation}{section}

% ----------------------------------------------------------------------------------------------------------------------------------------------------------------------
%								visibility calculation for modelling optical mode mismatch
% ----------------------------------------------------------------------------------------------------------------------------------------------------------------------
\section{Optimal state generation circuit}\label{appnA}

Consider a system comprised of linear optical components such as beam splitters, waveplates etc.  Assume $M$ modes entering this network undergo a unitary transformation described by
\beq   \label{U:LOS}  %Linear Optical System
\hat{\cal U}(G) = {\rm exp}\[- i \,\hat{\bf a}\dg G \,\hat{\bf a}\],
\eeq
where $\hat{\bf a} = \(\hat{a}_1, \hat{a}_2, \cdots, \hat{a}_M\)^\top$ is a vector of annihilation operators 
and $G$ is an Hermitian matrix. Output modes are obtained in the Heisenberg picture according to
\beq   \label{UaU:LOS}
\hat{\cal U}\dg(G) \,\hat{\bf a}\; \hat{\cal U}(G) = S(G)\, \hat{\bf a},
\eeq
where $S(G) = {\rm exp}\[{- i G}\]$ is the matrix representation of a unitary transformation induced on $\hat{\bf a}$. Let us now assume that the input state can be expressed by applying a function $f$ of incoming modes into the system on the total vacuum state $\ket{\bf 0} = \ket{{\mathbb 0}_1, {\mathbb 0}_2, \cdots, {\mathbb 0}_M}$ such that
\beq   \label{psi:in:fnot}   
\ket{\psi_{\rm in}} = f\(\hat{a}_1\dg, \hat{a}_2\dg, \cdots, \hat{a}_M\dg\) \ket{\bf 0}.
\eeq
The output state is simply obtained by unitarily evolving the above state, that is, $\ket{\psi_{\rm out}} = \hat{\cal U}(G) \ket{\psi_{\rm in}}$. It is straightforward to expand out this last relation using \erf{UaU:LOS} and the fact that $\hat{\cal U}\dg(G) = \hat{\cal U}(-G)$ and $S\dg(G) = S(-G)$ to get the following form
\beq   \label{psi:out:fnot}
\ket{\psi_{\rm out}} = f\(\hat{\bf a}\dg S_{:1}, \hat{\bf a}\dg S_{:2}, \cdots, \hat{\bf a}\dg S_{:M}\) \ket{\bf 0},
\eeq
where $S_{:m}$ denotes the $m$-th column of the matrix $S$ and $\hat{\bf a}\dg = (\hat{a}_1\dg, \hat{a}_2\dg, \cdots, \hat{a}_M\dg)$. Finding the matrix $S$ is thus central in our calculations to determine the output state of a quantum circuit composed of linear optical elements.

Let us now investigate how the two entangling gates produce the optimal state. With the input state $\ket{\psi_{bc}}$ same as in \erf{psi:bc:gen} and $\ket{\psi_a} = \ket{H}$ so that
\bqa \label{psi:in:NCN}
\bket{\psi_{\rm in}^{\rm NCN}} &=& f\(\hat v_a\dg, \hat{a}_V\dg, \hat{a}_H\dg, \hat{b}_H\dg, \hat{b}_V\dg, \hat{c}_V\dg, \hat{c}_H\dg, \hat v_c\dg\) \ket{\bf 0}, \label{psi:in:NCN1} \\
&=& \Big( \alpha_0 \, \hat{a}_H\dg \hat{b}_H\dg \hat{c}_H\dg + \alpha_1 \, \hat{a}_H\dg \hat{b}_H\dg \hat{c}_V\dg + \nn \\
&& \hspace{6pt} \alpha_2 \, \hat{a}_H\dg \hat{b}_V\dg \hat{c}_H\dg +  \alpha_3 \, \hat{a}_H\dg \hat{b}_V\dg \hat{c}_V\dg \Big) \,\ket{\bf 0}. \label{psi:in:NCN2}
\eqa
Note that the vacuum modes may be removed from the calculations as they are relevant for the second CNOT gate operation. The next step involves constructing the corresponding $S^{\rm NCN}$ matrix by multiplying matrices describing different optical components affecting appropriate modes; here four BSs and one mode swapping operation while leaving modes $c_V$ and $c_H$ unchanged. This process leads to 
\begin{widetext}
	\beq  \label{S:CNOT:log}
	S^{\rm NCN} = 
	\begin{pmatrix}
		1 &  0 & 0 & 0 & 0& 0 & 0 & 0 \\
		0 & {\eta_1} & \sqrt{\eta_1 (1-\eta_1)} & -\sqrt{\eta_1 (1-\eta_1)} &  1-\eta_1 & 0 & 0 & 0\\
		0 &\sqrt{\eta_1 (1-\eta_1)} & 1-\eta_1  & \eta_1 & -\sqrt{\eta_1 (1-\eta_1)} & 0 & 0 & 0\\
		0 &-\sqrt{\eta_1 (1-\eta_1)} & \eta_1 &  1-\eta_1  & \sqrt{\eta_1 (1-\eta_1)} & 0 & 0 & 0 \\
		0 &{1-\eta_1} & -\sqrt{\eta_1 (1-\eta_1)} & \sqrt{\eta_1 (1-\eta_1)} & 1-\eta_1& 0 & 0 & 0\\
		0 & 0 & 0 & 0 & 0& 1 & 0 & 0 \\
		0 & 0 & 0 & 0 & 0& 0 & 1 & 0\\
		0 &  0 & 0 & 0 & 0& 0 & 0 & 1
	\end{pmatrix},
	\eeq 
\end{widetext}
where the same mode ordering as the one appearing in the argument of function $f$ in \erf{psi:in:NCN} is employed for the matrix representation.  Now using \erf{psi:out:fnot} the total output state after applying the first CNOT gate becomes
\bqa \label{psi:out:NCN}
\bket{\psi_{\rm out}^{\rm NCN}} &=& f\( \hat{\bf a}\dg S_{:1}^{\rm NCN} , \cdots, \hat{\bf a}\dg S_{:8}^{\rm NCN} \) \ket{\bf 0} \\
&=& \Big[\alpha_0 \(\hat{\bf a}\dg S_{:3}^{\rm NCN} \) \( \hat{\bf a}\dg S_{:4}^{\rm NCN}\) \( \hat{\bf a}\dg S_{:7}^{\rm NCN}\) + \nn \\
&&  \alpha_1 \(\hat{\bf a}\dg S_{:3}^{\rm NCN} \)\( \hat{\bf a}\dg S_{:4}^{\rm NCN}\)  \( \hat{\bf a}\dg S_{:6}^{\rm NCN}\) + \nn \\
&& \alpha_2 \(\hat{\bf a}\dg S_{:3}^{\rm NCN} \)\( \hat{\bf a}\dg S_{:5}^{\rm NCN}\)  \( \hat{\bf a}\dg S_{:7}^{\rm NCN}\) + \nn \\
&& \alpha_3 \(\hat{\bf a}\dg S_{:3}^{\rm NCN} \)\( \hat{\bf a}\dg S_{:5}^{\rm NCN}\)  \( \hat{\bf a}\dg S_{:6}^{\rm NCN}\) \Big] \, \ket{\bf 0} \nn \\
&=& \frac{1}{\sqrt{2}} (\ket{\psi_1} + \ket{\psi_d}), \label{app:psi:out:ab}
\eqa
where
\beq
\hat{\bf a}\dg = \(\hat v_a\dg, \hat{a}_V\dg, \hat{a}_H\dg, \hat{b}_H\dg, \hat{b}_V\dg, \hat{c}_V\dg, \hat{c}_H\dg, \hat v_c\dg\),
\eeq
and $\ket{\psi_d}$ represents a superposition of states with more than one photon in either of qubit's modes.

The final output state from the circuit diagram shown in \frf{fig:circ:imper} calculated by setting the input state entering the CN gate as 
\bqa
\ket{\psi_{\rm in}^{\rm CN}} &=& \ket{\psi_{\rm out}^{\rm NCN}} \\
&=& g\(\hat v_a\dg, \hat{a}_V\dg, \hat{a}_H\dg, \hat{b}_H\dg, \hat{b}_V\dg, \hat{c}_V\dg, \hat{c}_H\dg, \hat v_c\dg\) \ket{\bf 0},
\eqa
and applying the corresponding matrix $S^{\rm CN}$ which is  
\begin{widetext}
	\beq  \label{S:CNOT:uni}
	S^{\rm CN} = 
	\begin{pmatrix}
		-\sqrt{\eta_2} & \sqrt{(1-\eta_2)} & 0 & 0 & 0 & 0 & 0 & 0 \\
		\sqrt{1-\eta_1} & \sqrt{\eta_2}  & 0 & 0 & 0 & 0 & 0 & 0\\
		0 & 0 & -\sqrt{\eta_2}  & -\sqrt{\eta_1 (1-\eta_1)} & \sqrt{(1-\eta_1)(1-\eta_2)} & 0 & 0 & 0  \\
		0 & 0 & 0 & 1 & 0& 0 & 0 & 0 \\
		0 & 0 & 0 & 0 & 1& 0 & 0 & 0 \\
		0 & 0 & \sqrt{\eta_1 (1-\eta_2)} & (1-2\eta_1) \sqrt{\eta_2} & 2\sqrt{\eta_1 \eta_2 (1-\eta_1)} & \sqrt{(1-\eta_1)(1-\eta_2)} & 0 & 0\\
		0 & 0 & \sqrt{(1-\eta_1)(1-\eta_2)} & 2\sqrt{\eta_1\eta_2 (1-\eta_1)} & (1-2\eta_1) \sqrt{\eta_2} & -\sqrt{\eta_1 (1-\eta_2)} & 0 & 0\\
		0 & 0 & 0 & \sqrt{(1-\eta_1)(1-\eta_2)} & \sqrt{\eta_1 (1-\eta_2)} & -\sqrt{\eta_2} & 0 & 0
	\end{pmatrix},
	\eeq 
\end{widetext}
to obtain
\bqa 
\bket{\psi_{\rm out}^{\rm CN}} &=& g\( \hat{\bf a}\dg S_{:1}^{\rm CN} , \cdots, \hat{\bf a}\dg S_{:8}^{\rm CN} \) \ket{\bf 0}, \label{psi:out:CNOT} \\ 
&=& \frac{1}{\sqrt{18}} \, \ket{\psi_{\rm opt}} + \ket{\chi_d}. \label{psi:out:ac}
\eqa
Here function $g$ can be found via \erf{app:psi:out:ab} and  $\ket{\chi_d}$ contains states with at least two photons in a single polarization or vacuum mode
\section{Phase measurement circuit}\label{appnB}
Consider the Heisenberg-limited phase estimation scheme with $K+1=3$ photons. Let us assume the input state is represented by $\rho_{\rm in}$. Recall the circuit shown in \frf{fig:3photab}, the state $\rho^{(K)} \in {\mathbb B}^{2^{K+1}}$ of the system before the first $X$-measurement on the $K$-th photon is
\beq    \label{rho:Kth}
\rho^{(K)} = \({\hat U}^{2^K}\bigotimes_{k=1}^K \hat{I}\) \rho_{\rm in}\,\({\hat U}^{2^K}\bigotimes_{k=1}^K \hat{I}\)\dg,
\eeq
where 
\beq  \label{phase:mat}
\hat U^m = 
\begin{pmatrix}
	1 & 0 \\
	0 & e^{im\phi}
\end{pmatrix}, \qquad
\hat{I} = 
\begin{pmatrix}
	1 & 0 \\
	0 & 1
\end{pmatrix}. 
\eeq 
Here $\phi$ is the unknown phase shift. The result of the $X$-measurement on the $K$-th photon is determined by the following measurement operators
\beq \label{mmnt:opt:Kth}
\hat{\cal M}_{r}^{(K)} = \hat{\Pi}_{r} \bigotimes_{k=1}^K \hat{I},  %\qquad {\rm for} \quad l = D, A.
\eeq 
where $r \in \{+,-\}$ is a measurement result, and $\hat{\Pi}_{r} = \op{r}{r}$ is the projection operator onto the $X$ basis of the $K$-th photon. Therefore, the probability of finding the $K$-th photon in one of the $X$ bases is 
\beq
\wp_{r}^{(K)} = \tr{\rho^{(K)} {\hat{\cal M}_{r}^{(K)}}\dg {\hat{\cal M}_{r}^{(K)}}} = \tr{\rho^{(K)} \hat{\cal M}_{r}^{(K)}}.
\eeq
Stochastic numerical simulations determine if the $K$-th photon is found in either of $\ket{\pm}$. Depending on the result of this measurement, the conditional system state $\rho_r^{(K)}$ after the measurement on the $K$-th photon is found via quantum measurement theory given by~\cite{Car08,WisMil10} 
\beq
\rho_r^{(K)} = \Big[\hat{\cal M}_{r}^{(K)} \rho^{(K)} {\hat{\cal M}_{r}^{(K)}}\dg \Big]/\wp_{r}^{(K)} = \hat{\Pi}_r \otimes \rho_r^{(K-1)},
\eeq
where $ \rho_r^{(K-1)}  \in {\mathbb B}^{2^{K}}$ is the reduced state matrix of the other remaining $K$ photons. 
The next step of the protocol includes some control operation depending on the result of the previous read out. That is, the measurement result $r = +(-)$ corresponds to feedback ``OFF(ON)" setting. Thus, in the reduced-dimension Hilbert space of the system, the state matrix before the measurement on the $(K-1)$-th photon when the control operation is ON can be written as 
\beq
\rho^{(K-1)} = V^{(K-1)} \, \rho_r^{(K-1)} \; {V^{(K-1)}}\dg,
\eeq
where 
\bqa  
V^{(K-1)} &\equiv& \[{\hat U}^{2^{K-1}}R\(\frac{\pi}{2}\)\] \bigotimes_{k=2}^K R\(\frac{\pi}{2^k}\), \\[4pt] 
R{(\theta)} &=& 
\begin{pmatrix} 
	e^{i\theta/2} & 0 \\
	0 & e^{-i\theta/2}
\end{pmatrix}, \label{FB:phase}
\eqa 
and if the control operation is OFF,  the state matrix is 
\beq   \label{FB:off:K-1}
\rho^{(K-1)} = \({\hat U}^{2^{K-1}}\bigotimes_{k=1}^K \hat{I}\) \rho_r^{(K-1)} \({\hat U}^{2^{K-1}}\bigotimes_{k=1}^K \hat{I}\)\dg.
\eeq
Using the same procedure the measurement on the $(K-1)$-th photon is described. In other words, by changing $K\to K-1$ we can recall \erfs{mmnt:opt:Kth}{FB:off:K-1} to obtain the measurement outcome and the reduced state $ \rho_r^{(K-2)}  \in {\mathbb B}^{2^{K-1}}$ of the system. These steps are repeated for each photon until the zeroth one, for which, the measurement operator is simply the projector $\hat{\cal M}_{r}^{(0)} = \hat{\Pi}_{r}$.

\section{Calculations of the shot-noise-limit} \label{appnSNL}
Consider the interferometer illustrated in \frf{fig:GMZI} with $p=1$.  
In ideal experimental scenario in which a single photon is incident into one arm of the interferometer, the probability of detecting a photon in either of the outputs ports is
\beq
p(u|\phi,\theta) = \frac{1}{2} \left[1+u\,{\rm cos}(\phi-\theta) \right],
\eeq 
where $u \in \{-1,1\}$ is the measurement result. Assuming that $m$ measurement outcomes are obtained, one can define a vector ${\bf u}_m = \{u_1, u_2, \cdots, u_m \}$ in
which each $u_\ell$ is defined as above. Therefore, the probability for the series of measurement results is
\beq
p({\bf u}_m|\phi,{\boldsymbol\theta}) = \frac{1}{2} \left[1+u\,{\rm cos}(\phi-\theta) \right],
\eeq 
where ${\boldsymbol\theta}=\{\theta_1,\theta_2,\cdots,\theta_m\}$ are the measurement angles for each photon. The Holevo variance in the phase estimate is $V_H = \mu^{-2} -1$ where $\mu = |\langle e^{i \phi} \rangle|$ and using the above probability distribution can be expressed as
\beq
\mu = \frac{1}{2 \pi} \sum_{{\bf u}_m} \left| \int d\phi \, e^{i \phi}\, p({\bf u}_m|\phi,{\boldsymbol\theta}) \right|.
\eeq 
Calculating this integral for a small number of resources ($N=3,7$), one can obtain the SNL for Holevo variance. 
While these solutions are not unique, we find that setting ${\boldsymbol\theta}=\{0,0,\pi/2\}$ Rad would achieve $V_{\rm SNL} = 0.655845...$ for an $N=3$ measurement, while the measurement angles ${\boldsymbol\theta}=\{0, 0, 2.31099, 1.32133, 1.32133, 0.843774, -0.830605\}$ Rad would achieve $V_{\rm SNL} = 0.232688...$ for an $N=7$ measurement.

\section{HOM visibility and spatial optical mode mismatch} \label{appnC}
Recall \erf{pcoin:smm}, the probability for observing a coincidence photon detection is
\bqa   \label{pcoin:smm:appn}
p_{\rm coin} &=& \an{\hat{n}_{a_D} \hat{n}_{b_D}} \\
&=& \bra{\bf 01}(\hat{n}_{\hat{v}_2^{\rm o}} + \hat{n}_{\hat{a}^{\rm o}} + \hat{n}_{\hat{v}_1^{\rm o}}) (\hat{n}_{\hat{v}_4^{\rm o}} + \hat{n}_{\hat{b}^{\rm o}} + \hat{n}_{\hat{v}_3^{\rm o}})  \ket{\bf 0 1},  \nonumber 
\eqa
where 
\beq 
\ket{\bf 01} \equiv \ket{{\mathbb 0}}_{\hat{v}_1^{\rm o}} \otimes \ket{{\mathbb 0}}_{\hat{v}_2^{\rm o}} \otimes \ket{{\mathbb 0}}_{\hat{v}_3^{\rm o}} \otimes \ket{{\mathbb 0}}_{\hat{v}_4^{\rm o}} \otimes \ket{{\mathbb 1}}_{\hat{a}^{\rm o}} \otimes \ket{{\mathbb 1}}_{\hat{b}^{\rm o}}.
\eeq
We use superscript ``o" to denote output modes corresponding to the relevant input modes. There are 9 terms that should be calculated separately to obtain a relation for the probability as a function of mode mismatch parameters. It is straightforward to show that the output modes are given by the following equations:
\begin{subequations}
\bqa
\hat{v}_1^{\rm o} &=& \sqrt{(1-\eta)(1-\xi_2)} \,\hat{b} - \sqrt{\xi_2 (1-\eta)} \,\hat{v}_4 -  \sqrt{\eta}\, \hat{v}_1, \nn \\
\hat{a}^{\rm o} &=& \sqrt{(1-\eta)(1-\xi_2)}\,\hat{v}_4 + \sqrt{\xi_2 (1-\eta)} \, \hat{b}  - \nn \\
&&  \sqrt{\eta(1-\xi_1)} \,\hat{v}_2- \sqrt{\eta \xi_1}\, \hat{a}, \nn \\
\hat{v}_2^{\rm o} &=& \sqrt{1-\eta} \,\hat{v}_3 - \sqrt{\eta(1-\xi_1)} \,\hat{a} + \sqrt{\eta \xi_1} \,\hat{v}_2, \nn \\ 
\hat{v}_3^{\rm o} &=& \sqrt{(1-\eta)(1-\xi_1)} \,\hat{a} - \sqrt{\xi_1 (1-\eta)} \hat{v}_2 + \sqrt{\eta} \hat{v}_3, \nn \\
\hat{b}^{\rm o} &=&  \sqrt{(1-\eta)(1-\xi_1)}\,\hat{v}_2 + \sqrt{\xi_1 (1-\eta)} \, \hat{a} + \nn \\
&& \sqrt{\eta(1-\xi_2)} \,\hat{v}_4 - \sqrt{\eta \xi_2}\, \hat{b}, \nn \\
\hat{v}_4^{\rm o} &=&  \sqrt{(1-\eta)} \, \hat{v}_1 + \sqrt{\eta (1-\xi_2)} \, \hat{b} -\sqrt{\eta \xi_2} \hat{v}_4. \nn
\eqa 
\end{subequations}
Now by using the these equations and the commutation relations for creation and annihilation operators of each mode, all terms in \erf{pcoin:smm:appn} can be calculated in a straightforward manner. For instance, the first term is equal to
\bqa
&&\bra{\bf 01} \hat{n}_{\hat{v}_2^{\rm o}} \hat{n}_{\hat{v}_4^{\rm o}} \ket{\bf 01} =  \bra{\bf 01} \;\hat{v}_2^{\rm o \dagger} \hat{v}_2^{\rm o} \hat{v}_4^{\rm o \dagger} \hat{v}_4^{\rm o} \; \ket{\bf 01}, \nonumber \\
&& \hspace{10pt} =\bra{\bf 01} \[  \sqrt{1-\eta} \,\hat{v}_3\dg - \sqrt{\eta(1-\xi_1)} \,\hat{a}\dg + \sqrt{\eta \xi_1} \,\hat{v}_2\dg\] \times \nonumber \\
&&\hspace{40.5pt} \[ \sqrt{1-\eta} \,\hat{v}_3 - \sqrt{\eta(1-\xi_1)} \,\hat{a} + \sqrt{\eta \xi_1} \,\hat{v}_2\] \times \nonumber \\
&& \hspace{40.5pt} \[  \sqrt{(1-\eta)} \, \hat{v}_1\dg + \sqrt{\eta (1-\xi_2)} \, \hat{b}\dg -\sqrt{\eta \xi_2} \hat{v}_4\dg \] \times \nonumber \\
&&\hspace{40.5pt} \[ \sqrt{(1-\eta)} \, \hat{v}_1 + \sqrt{\eta (1-\xi_2)} \, \hat{b} -\sqrt{\eta \xi_2} \hat{v}_4\]  \ket{\bf 01}, \nonumber \\
&&\hspace{10pt} =\[\eta^2 (1-\xi_1)(1-\xi_2)\]  \bra{\bf 1} \hat{a}\dg \hat{a} \,\hat{b}\dg \hat{b}  \ket{\bf 1}, \nonumber \\
&&\hspace{10pt} =\eta^2 (1-\xi_1)(1-\xi_2),
\eqa
where $\ket{\bf 1} = \ket{1}_{\hat{a}^{\rm o}} \otimes \ket{1}_{\hat{b}^{\rm o}}$ and in the second last line we just kept nonzero terms. Applying the same procedure to the rest of terms in \erf{pcoin:smm:appn} gives
\begin{subequations}
\bqa
\bra{\bf 01} \hat{n}_{\hat{v}_2^{\rm o}} \hat{n}_{\hat{b}^{\rm o}} \ket{\bf 0 1} & = & \eta^2\, \xi_2 \,(1-\xi_1), \\
\bra{\bf 01} \hat{n}_{\hat{v}_2^{\rm o}} \hat{n}_{\hat{v}_3^{\rm o}} \ket{\bf 0 1} & = & 0, \\
\bra{\bf 01} \hat{n}_{\hat{a}^{\rm o}} \hat{n}_{\hat{v}_4^{\rm o}} \ket{\bf 0 1} & = & \eta^2\, \xi_1 \,(1-\xi_2), \\
\bra{\bf 01} \hat{n}_{\hat{a}^{\rm o}} \hat{n}_{\hat{b}^{\rm o}} \ket{\bf 0 1} & = & \xi_1 \xi_2 \[1-4\eta(1-\eta)\], \\
\bra{\bf 01} \hat{n}_{\hat{a}^{\rm o}} \hat{n}_{\hat{v}_3^{\rm o}}  \ket{\bf 0 1} & = & (1-\eta)^2\, \xi_2 \,(1-\xi_1), \\
\bra{\bf 01} \hat{n}_{\hat{v}_1^{\rm o}}\hat{n}_{\hat{v}_4^{\rm o}} \ket{\bf 0 1} & = & 0, \\
\bra{\bf 01} \hat{n}_{\hat{v}_1^{\rm o}} \hat{n}_{\hat{b}^{\rm o}} \ket{\bf 0 1} & = & (1-\eta)^2\, \xi_1 \,(1-\xi_2), \\
\bra{\bf 01} \hat{n}_{\hat{v}_1^{\rm o}}\hat{n}_{\hat{v}_3^{\rm o}} \ket{\bf 0 1} & = & (1-\eta)^2\, (1-\xi_1) (1-\xi_2).
\eqa
\end{subequations}
Summing up all of these terms gives \erf{pcoin:smm}
\beq  
p_{\rm coin} = \an{\hat{n}_{a_D} \hat{n}_{b_D}} = \frac{1}{2}\(1-\xi_1\xi_2\).
\eeq
The dip visibility can easily be worked out now once we have the probability of coincidence detection.

\bibliography{references}

\end{document}